\documentclass[11pt]{article}
\pdfoutput=1
\usepackage{jheppub}
\usepackage[T1]{fontenc}
\usepackage{cancel,tensor,enumitem,fix-cm}
\usepackage{epsfig}
\usepackage{graphicx}
\usepackage[english]{babel}
\usepackage{amsmath}
\usepackage{textcomp}
\usepackage{multirow}
\usepackage{booktabs}
\usepackage{tikz}
\usepackage{overpic}
\newcommand{\es}[2] {\begin{equation} \label{#1} \begin{split} #2 \end{split} \end{equation}}
\newcommand\ov{\over}
\def\le{\left}
\def\ri{\right}
\def\pa{\partial}
\newcommand\abs[1]{\ensuremath{\left\lvert{#1}\right\rvert}}

\def\one{{\,\hbox{1\kern-.8mm l}}}
\newcommand{\Dslash}{\not{\hbox{\kern-4pt $D$}}}
\newcommand{\pdslash}{\not{\hbox{\kern-2pt $\partial$}}}

\newcommand{\Comment}[1]{{}}

\newcommand{\hN}{\hat{N}}

\def\IZ{{\mathbb Z}}

\def\IR{{\mathbb R}}

\def\({\left(}
\def\){\right)}

\newcommand{\bc}{\begin{center}}
\newcommand{\ec}{\end{center}}
\newcommand{\ba}{\begin{array}}
\newcommand{\ea}{\end{array}}
\newcommand{\bea}{\begin{eqnarray}}
\newcommand{\eea}{\end{eqnarray}}
\newcommand{\bmx}{\begin{pmatrix}}
\newcommand{\emx}{\end{pmatrix}}

\newcommand{\be}{\begin{equation}}
\newcommand{\ee}{\end{equation}}

\newcommand{\om}{\omega}

\newcommand{\vw}{{\vec{w}}}

\newcommand{\hn}{{\hat{n}}}

\def\IB{\relax{\rm I\kern-.18em B}}
\def\IC{{\relax\hbox{\kern.3em{\cmss I}$\kern-.4em{\rm C}$}}}
\def\ID{\relax{\rm I\kern-.18em D}}
\def\IE{\relax{\rm I\kern-.18em E}}
\def\IF{\relax{\rm I\kern-.18em F}}
\def\II{\relax{\rm I\kern-.18em I}}
\def\IZ{\relax{\sf Z\kern-.35em Z}}
\def\Id{\relax{1\kern-.32em 1}}
\def\IG{\relax\hbox{$\inbar\kern-.3em{\rm G}$}}
\def\IR{\relax{\rm I\kern-.18em R}}

\normalsize

\subheader{\begin{flushright}YITP-19-36
\end{flushright}}

\title{Bit Threads and the Membrane Theory of Entanglement Dynamics}

\author[a]{Cesar A. Ag\'on,}
\author[b]{and M\'ark Mezei}
\affiliation[a]{C.~N. Yang Institute for Theoretical Physics, State University of New York\\
Stony Brook, NY 11794, USA}
\affiliation[b]{Simons Center for Geometry and Physics, SUNY, \\
Stony Brook, NY 11794, USA}
\emailAdd{cesar.agon@stonybrook.edu}
\emailAdd{mmezei@scgp.stonybrook.edu}

\abstract{
Recently,  an effective {\it membrane theory} was proposed that describes the ``hydrodynamic'' regime of the entanglement dynamics for general chaotic systems. 
Motivated by the new {\it bit threads} formulation of  holographic entanglement entropy, given in terms of a convex optimization problem based on flow maximization, or equivalently tight packing of  bit threads, we reformulate the membrane theory as a max flow problem by proving a max flow-min cut theorem. 
In the context of holography, we explain the relation between the max flow program dual to the membrane theory and the max flow program dual to the holographic surface extremization prescription by providing an explicit map from the membrane to the bulk, and derive the former from the latter in the  ``hydrodynamic'' regime without reference to minimal surfaces or membranes.
}

\begin{document}
\maketitle
\flushbottom

\section{Introduction}

Entanglement has been playing an important role in recent studies of holography \cite{Ryu:2006bv,Ryu:2006ef,Hubeny:2007xt,VanRaamsdonk:2010pw,Hartman,Maldacena:2013xja,Liu:2013iza,Liu:2013qca} and the thermalizing out-of-equilibrium dynamics \cite{bardarson2012unbounded,serbyn2013universal,Kim:2013bc,kaufman2016quantum}. It is expected that in the large subregion, long time limit, the dynamics of entanglement entropy simplifies. Indeed, in this ``hydrodynamic'' limit recently a simple effective theory, the membrane theory was proposed in \cite{Jonay:2018yei} and elaborated upon in \cite{Mezei:2018jco}, based on the earlier results of  \cite{Nahum:2016muy,Mezei:2016wfz,Mezei:2016zxg}. In holography the HRT prescription \cite{Ryu:2006bv,Ryu:2006ef,Hubeny:2007xt} gives an elegant geometric prescription for computing the  entanglement entropy of field theories with a holographic dual. It has been recently reformulated in the language of bit threads \cite{Freedman:2016zud}. In this paper we investigate the interplay of these two recent developments.  For recent work on bit threads see for example \cite{Hubeny:2018bri,Cui:2018dyq,Agon:2018lwq,Harper:2018sdd,Bao:2019wcf,Du:2019emy,Harper:2019lff}. We summarize our results and give the outline of the paper below.

To set the stage for our discussion, we give the statement of the bit thread reformulation of the HRT prescription. The reformulation is simplest to state in the static RT case \cite{Ryu:2006bv,Ryu:2006ef}, where the entropy is computed in two equivalent ways:
\es{BitThreadReform}{
S[A]={1\ov 4G_N}\,\min_{m\sim A}\text{area}(m) \quad  \iff \quad S[A]&={1\ov 4G_N}\, \max_{w^\mu\in {\cal F}}\int_A \sqrt{h} \, n_\mu w^\mu\,, \\
 {\cal F}&\equiv\{w^\mu\, \vert\, \nabla_\mu w^\mu=0,\, \abs{w^\mu}\leq 1\}\,.
}
The LHS is the RT prescription: we are instructed to find the minimal area surface $m$ that is homologous to the boundary region $A$ whose entropy we are interested in. The RHS is the bit thread reformulation: we are interested in finding a divergence free vector field $w^\mu$ with bounded norm that maximizes the flux through $A$\footnote{These divergence free vector fields $w^\mu$ are Hodge dual to closed forms called calibrations \cite{Harvey:1982xk}. Recently, the authors of \cite{Bakhmatov:2017ihw} considered a similar reformulation of the RT formula in terms of calibrations. }. (On the RHS $n_\mu$ is the unit normal covector field of the AdS boundary.)  We can equivalently think of the vector field  $w^\mu$ as bit threads of Planck width that connect the subsystem $A$ to the rest of the system: they represent EPR pairs shared between $A$ and the rest of the system and their number gives the entropy. The equivalence of the results of the two optimization problems is a consequence (of the Riemannian version) of the max flow-min cut theorem of network theory. Intuitively, since the vector field is divergence free, we can evaluate its flux through any region homologous to $A$. There is a bottleneck for increasing the flux, at the minimal surface $m$ the vector field is $w^\mu=n^\mu_{(m)}$, where $n^\mu_{(m)}$ is the unit normal covector field of $m$ as a hypersurface, and the two sides agree.

\begin{center}
\begin{figure}[!h]
\centering
\includegraphics[scale=0.6]{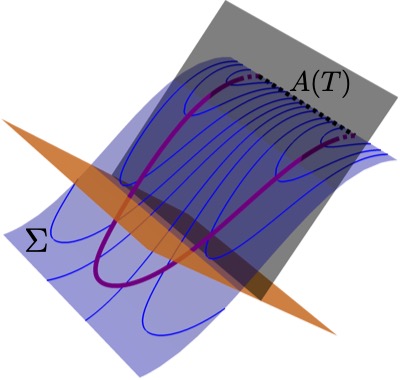}
\caption{In this figure we illustrate the bit thread reformulation of the HRT prescription \eqref{HRTBitThreadReform}. We focus on the spacetime  of a black hole forming from collapse which is dual to a field theory quench: 
on the figure, the orange and the black planes represent the infalling matter shell and the black brane horizon respectively. The  three directions on the diagram are time (vertical), the bulk radial and the field theory spatial directions. We want to understand the entanglement entropy $S[A(T)]$ of subregion $A(T)$ in the field theory, which is computed by the area of the HRT surface (solid purple line)  anchored on $\partial A(T)$.
We use Kruskal coordinates, hence the AdS boundary is infinitely far away, which we indicate by using dashed lines.   
 We drew a transparent blue Cauchy slice $\Sigma$ through the bulk that contains the HRT surface. The  blue lines represents the flow lines of $w^\mu$, the bit threads. They are floppy away from the HRT surface, but pass through it perpendicularly, and they are maximally packed at this bottleneck. Their flux through $A(T)$ also computes $S[A(T)]$. The membrane theory describes the part of the HRT surface that lies between the matter shell and the horizon. This portion of the HRT surface determines the leading extensive piece of the entropy. The membrane bit threads approximate the flow in the same region, as we explain below. See Fig.~\ref{fig:Penrose} for a conventional Penrose diagram, as well as discussion around there for more explanation about the spacetime geometry. 
\label{fig:hrt}}
\end{figure}
\end{center}

We can extend the reformulation \eqref{BitThreadReform} to the time dependent HRT case, by using the maximin reformulation of HRT \cite{Wall:2012uf}: we fix a Cauchy surface $\Sigma$, perform area minimization on it, and then maximize over $\Sigma$. We can then straightforwardly use the bit thread reformulation for each Cauchy surface $\Sigma$, and then maximize over all Cauchy surfaces:
\es{HRTBitThreadReform}{
S[A(T)]={1\ov 4G_N}\,\max_{\Sigma  \supset \partial A(T)}\underset{\underset{m\subset\Sigma}{m\sim A(T)}}{\min}\text{area}(m) \quad  \iff \quad S[A(T)]&={1\ov 4G_N}\,\max_\Sigma \max_{w^\mu\in {\cal F}_\Sigma}\int_{A(T)} \sqrt{h} \, n_\mu w^\mu\,, \\
 {\cal F}_\Sigma&\equiv\{w^\mu\in \mathfrak{X}(\Sigma) \, \vert\, \nabla_\mu w^\mu=0,\, \abs{w^\mu}\leq 1\}\,,
}
where $\mathfrak{X}(\Sigma)$ is the space of vector fields on $\Sigma$. See Fig.~\ref{fig:hrt} for an illustration.  There also exists a covariant bit thread version of the correspondence \cite{Headrick:toappear}, but we will not use it in this paper.

In \cite{Mezei:2018jco}  the membrane theory was derived from the HRT prescription in the hydrodynamic limit; we review this derivation and provide an alternative one using the maximin reformulation  of the HRT prescription in Section~\ref{sec:membrane}. The membrane theory instructs us to compute the minimal membrane action
\es{MembraneTheoryIntro}{
S[A(T)]=s_\text{th}\min_{m\sim A(T)/(t=0)} \int_m \sqrt{h}\,\frac{\mathcal{E}(v)}{\sqrt{1+ v^2}}\,, \qquad v\equiv \frac{ (n\cdot \hat{t})}{\sqrt{1- (n\cdot \hat{t})^2}}\,,
}
where $s_{th}$ is the equilibrium thermal entropy density, $n$ is the unit normal to the membrane, $\hat{t}$ is the timelike unit vector, the notation $m\sim A/(t=0)$ means that $m$ is homologous to $A$ relative to the $t=0$ surface, and the membrane lives in Euclidean space. Since there is a fixed time coordinate in the formulation, we could as well rewrite the problem in Minkowski space, and this was the choice made in \cite{Mezei:2018jco} to emphasize that we are studying real time evolution.\footnote{If we choose Minkowski signature, \eqref{MembraneTheoryIntro} is modified to: 
\es{MembraneTheoryIntro-2}{
S[A(T)]=s_\text{th}\min_{m\sim A(T)/(t=0)} \int_m \sqrt{-h}\,\frac{\mathcal{E}(v)}{\sqrt{1-v^2}}\,, \qquad v\equiv \frac{ (n\cdot \hat{t})}{\sqrt{1+(n\cdot \hat{t})^2}}\,, 
}
where $h_{\alpha \beta}$ in the above equation is the  metric on $m$ induced from the Minkowski metric. Here, $n$ is the unit normal to the membrane in Minkowski signature.}  In this paper the Euclidean formulation is more convenient. 

The membrane theory applies in the dynamical setting, but mathematically it is more analogous to the static RT problem \eqref{BitThreadReform}. (In fact, we would recover the static RT problem \eqref{BitThreadReform} by setting ${\mathcal{E}(v)}=\sqrt{1+ v^2}$.)  It is natural to expect that similarly to the RT case, the membrane theory admits a bit thread reformulation. The main result of this paper, presented in  Section~\ref{2.2}, is the derivation of the reformulation:
\es{MemBitThreadReform}{
S[A(T)]&=s_{th}\, \max_{w^\mu\in {\cal F}}\int_{A(T)}  n_\mu w^\mu\,, \qquad {\cal F}\equiv\{w^\mu\, \vert\, \nabla_\mu w^\mu=0,\, \abs{\vec{w}}\leq H(w^t)\}\,,
}
where we used the decomposition $w^\mu=(w^t,\,\vec{w})$, and introduced the function $H(w^t)$ which is the Legendre transform of $-{\cal E}(v)$,\footnote{Note that for ${\mathcal{E}(v)}=\sqrt{1+ v^2}$ we recover the constraints  on $w^\mu$ given in \eqref{BitThreadReform}.}
\es{LegTrans}{
w^t=-{\cal E}'(v)\,, \qquad H(w^t)=\le({\cal E}(v)-v\,{\cal E}'(v)\ri)\Big\vert_{v=v(w^t)}\,.
}
See Fig. \ref{fig:membrane} for a pictorial representation of both, the minimal membrane, and the membrane bit threads.
\begin{center}
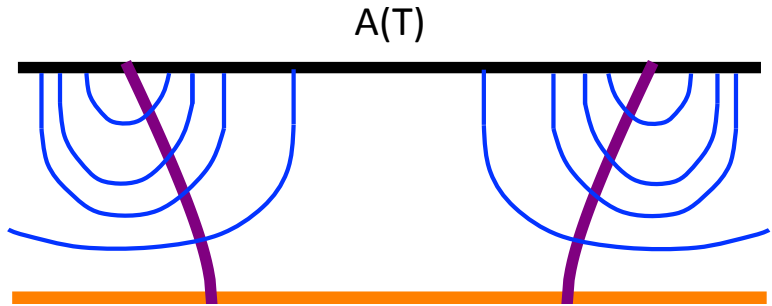
\begin{figure}[!h]
\centering
\includegraphics[scale=0.4]{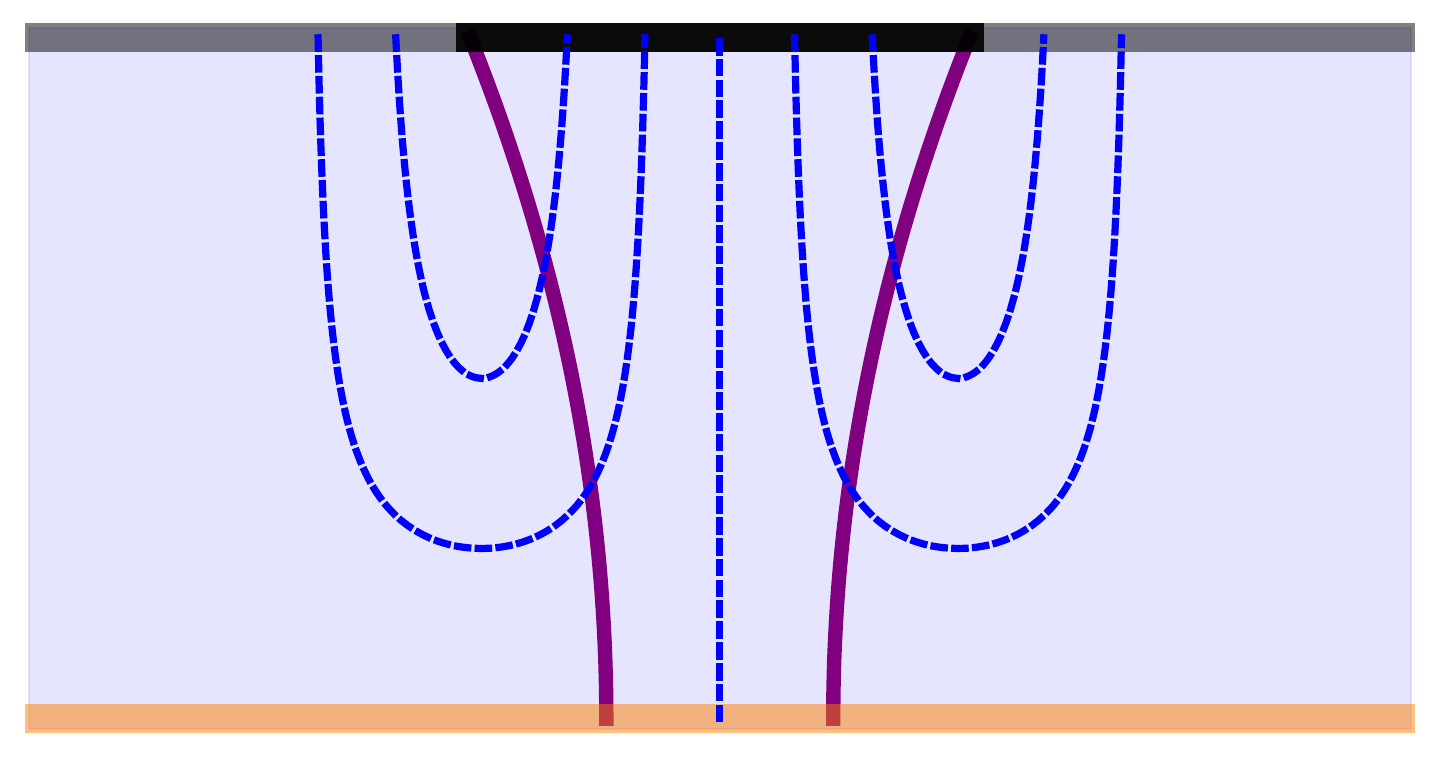}
 \setlength{\unitlength}{1cm}
  \setlength{\unitlength}{1cm}
\begin{picture}(0,0)
\put(-5.7,5.4){{$A(T)$}}
\end{picture}
\caption{In this figure we illustrate the bit thread reformulation of the membrane theory \eqref{MemBitThreadReform}. The initial state is drawn with orange, the time $T$ is represented by a gray line, and the purple membrane (anchored on $\pa A(T)$) is stretching between the two. 
(In this case $A(T)$ is a sphere, the vertical direction is time, while the horizontal is the field theory spatial radial coordinate.) The membrane bit threads are drawn with dashed blue lines. Note that some threads have a vertical section:
since for $\abs{v}\geq v_B$ we have $H(w^t)=0$ (see \eqref{tildeE(v)} for the explanation of this fact), it follows that for this case  $|\vec{w}|=0$, hence the threads point in the time direction. Comparing to Fig.~\ref{fig:hrt}, it is easy to visualize that the membrane picture is the projection of the bulk picture along constant infalling time: the infalling matter shell becomes the orange line, the intersection of the Cauchy surface with the horizon becomes the black line, while the appropriate part of the HRT surface and bulk bit threads become the membrane and the membrane bit threads. What is highly nontrivial is that we can write a theory in terms of the membrane or membrane bit threads alone.
We note that the membrane is a semi-analytic result from \cite{Mezei:2018jco}, while the bit threads are only sketches. 
\label{fig:membrane}}
\end{figure}
\end{center}

\begin{center}
\begin{figure}[!h]
\centering
\includegraphics[scale=0.6]{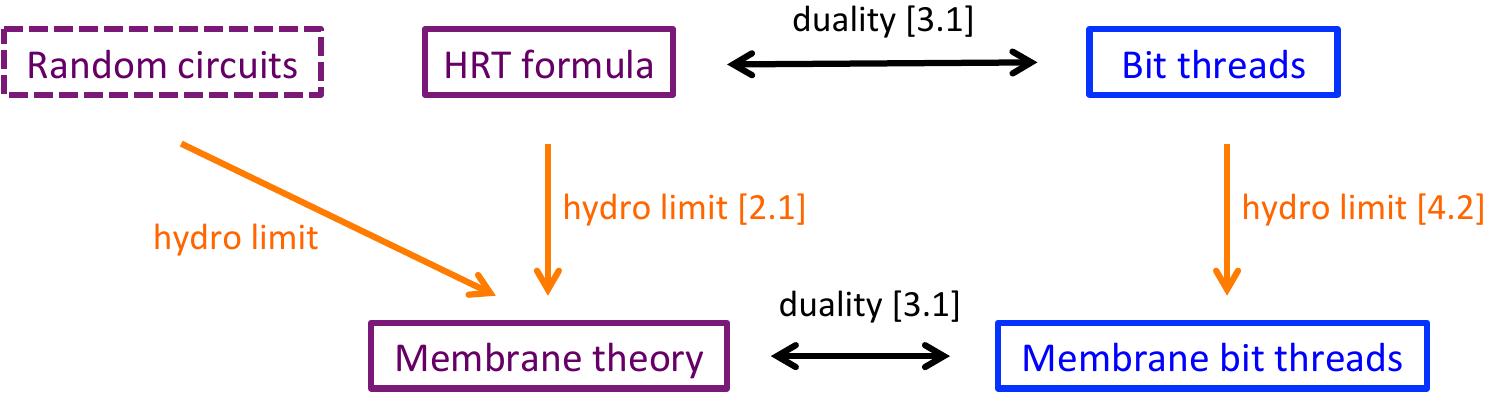}
\caption{On this flowchart we show the relations between the different theories that compute the entanglement entropy $S[A(T)]$. The numbers in brackets refer to the section where we discuss the particular relation.
\label{fig:flowchart}}
\end{figure}
\end{center}

It is natural to ask what does the minimal membrane and the corresponding bit threads have to do with the HRT surface and the bulk bit threads. The membrane is the projection of the HRT surface to the boundary of AdS along constant infalling time. In this setup the HRT surface can be recovered from the membrane using an explicit map. Using this same map the membrane bit threads can be pushed into the bulk, and we get bulk bit threads satisfying the constraints as we show in Section~\ref{sec:bulk}. There, we also show that starting from bulk bit threads, the RHS of \eqref{HRTBitThreadReform}, we can obtain the membrane bit thread description \eqref{MemBitThreadReform} in the hydrodynamic limit. This brings the logic of this paper to a full circle, as illustrated on the flowchart in Fig.~\ref{fig:flowchart}.

The membrane theory  is a  candidate to be the universal effective description of entanglement dynamics in the hydrodynamic regime. Even though currently there is  no general physical argument for its validity, we find it promising, since it can be derived from disparate physical systems: holographic gauge theories  \cite{Mezei:2018jco} and random quantum circuits \cite{Jonay:2018yei}. To gain more confidence in the theory it is a worthy enterprise to study its theoretical structure, and assess the feasibility of its properties.  
The main result of this paper is a dual description of the membrane theory in a flow or bit thread formulation, which makes no reference to the membrane. The two ends of the bit threads may be interpreted as the two members of EPR pairs that have been distilled from the entangled state. The floppiness (non-uniqueness) of the maximal bit thread configuration corresponds to the highly non-unique nature of the distillation process. The minimal membrane can be thought of as a coarse grained cut through the tensor network representation of the evolving wave function. The bit thread reformulation then raises the question what properties should the tensor network have, so that for any entanglement cut we can turn it into EPR pairs. Similarly, powerful many-party inequalities for entanglement entropy can be derived from the membrane theory \cite{Bao:2018wwd}. If the membrane theory is universally valid in all chaotic systems, all these properties should hold for the  wave function evolving from a short range initial state under chaotic time evolution in the hydrodynamic limit. It would be fascinating, if the flow reformulation developed in this paper would provide an avenue to proving the validity of this effective theory for entanglement dynamics.

\section{Hydrodynamic regime of entanglement dynamics}\label{sec:membrane}
 
\subsection{Membrane theory review\label{section2}}

We want to study the dynamics of entanglement in a homogenous highly excited out of equilibrium state. Such a state can be prepared experimentally using a quench. In a chaotic system the state will settle to thermal equilibrium with inverse temperature $\beta$. In the ``hydrodynamic regime'', where we study subsystems of characteristic size $R$ at time $T$, which obey
\es{Scaling}{
R,\, T\gg \beta\,,
}
we expect that the laws governing the dynamics of entanglement entropy simplify.\footnote{In a weakly coupled chaotic system, in \eqref{Scaling} we should replace $\beta$ with the local equilibration time $t_\text{loc}\sim \beta / g$, where $g$ is the coupling strength. We expect that the membrane theory applies to these systems as well at the longest distance and time scales. } The corresponding effective theory was proposed in \cite{Jonay:2018yei} based on results obtained from random quantum circuits \cite{Nahum:2016muy}, and the effective theory was derived from holography in \cite{Mezei:2018jco}. 

\begin{center}
\begin{figure}[!h]
\centering
\includegraphics[scale=0.8]{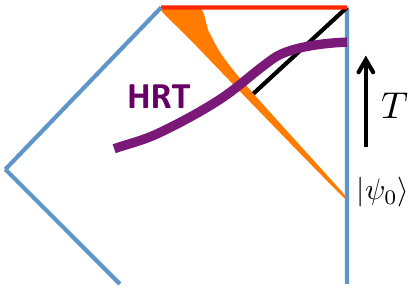}
\caption{Penrose diagram of a spacetime dual to a quench. Before the quench we have pure AdS, the infalling matter shell is colored orange, where the spacetime is strongly time dependent, and the spacetime subsequently settles to a static black brane.  The horizon is a diagonal black line, the singularity is a red line, the Poincare horizon and the AdS boundary are drawn by blue, while the HRT surface computing the entropy of half space is a purple curve.
\label{fig:Penrose}}
\end{figure}
\end{center}

We review the derivation briefly. The quench state $\vert{\psi_0}\rangle$ and its subsequent time evolution is dual to a spacetime represented in Fig~\ref{fig:Penrose}, where we also depict a typical HRT surface anchored on the boundary region $A(T)$. (For a three-dimensional Penrose diagram see Fig.~\ref{fig:hrt}.) Let us introduce a large bookkeeping parameter $\Lambda$ (which we set to $\Lambda=1$ at the end of the calculation, and replace $R,\, T\to \Lambda R,\,\Lambda T$. We want to describe the leading, extensive $\Lambda^{d-1}$ piece of the entropy. The first step of the derivation is to understand that we only get a contribution to this extensive piece from the behind the horizon region of the spacetime; the other parts of the HRT surface only contributes area law, $\Lambda^{d-2}$ pieces \cite{Mezei:2016zxg,Mezei:2018jco}.\footnote{This includes the divergent area law piece coming from the near boundary part of the geometry. Since the divergent piece is time independent, we can also discard it by considering the vacuum subtracted entanglement entropy $S[A(T)]\equiv S_{\vert{\psi_0}\rangle}[A(T)]-S_{\vert{0}\rangle}[A(T)]$.} Second, we can establish that the piece of the HRT surface connecting the boundary to the horizon is very simple to leading order: in Schwartzschild coordinates the HRT surface is just the entangling surface $\pa A(T)$ extended into the bulk along the bulk radial direction $z$ as a cylinder, and hence on the horizon its image is  $\pa A(T)$ \cite{Mezei:2016zxg}. We do not discuss these points further, as they are not central to what we set out to do in the rest of the paper.

Third, we analyze the HRT problem in the behind the horizon region, which we write in infalling coordinates as 
\bea\label{f-metric}
ds^2=\frac{1}{z^{2}}\left( -a(z)dt^{2} +d\vec{x}^{2}-\frac{2}{b(z)}dt dz  \right)\,,
\eea
and want to describe the codimension-2 HRT surface using the dependent coordinates $z(t,\Omega),\, \rho(t,\Omega)$, where $(\rho,\Omega)$ are  spherical coordinates on a fixed time slice in the boundary theory. Without loss of generality, we choose the horizon to be at $z_h=1$; we then have $a(1)=0$. The area functional is 
\es{AreaFunct}{
S[A(T)]&={1\ov 4G_N}\min_{m\sim A(T)}\int dt\, d\Omega \ {\rho^{d-2}\ov z^{d-1}}\sqrt{Q}\\
Q&\equiv \le[\dot\rho^2-\le(1+{(\pa_\Omega \rho)^2\ov \rho^2}\ri)a(z)\ri]+{2\ov b(z)}\le[{\dot\rho \,(\pa_\Omega \rho \cdot \pa_\Omega z)\ov \rho^2}-\le(1+{(\pa_\Omega \rho)^2\ov \rho^2}\ri)\dot z\ri]-{(\pa_\Omega z)^2\ov \rho^2 b(z)^2}\,.
}
Now we implement the scaling limit: we scale the field theory coordinates, but not the bulk radial coordinate $z$:
\es{CoordScaling}{
t\to\Lambda t\,, \quad \rho\to\Lambda \rho\,, \quad \Omega\to\Omega \,, \quad z\to z\,.
}
Plugging into \eqref{AreaFunct} we get an overall $\Lambda^{d-1}$ factor that corresponds to extensivity, while to leading order in $\Lambda$, only the first term in $Q$ remains. This term only depends on $z$ algebraically (compare to the full action), thus the equation of motion of $z$ is algebraic
\es{zEOM}{
a(z)-\frac{z a'(z)}{2(d-1)}= {\dot\rho^2\ov 1+{(\pa_\Omega \rho)^2\ov \rho^2}}\,.
}
Let us introduce some notation, we define:
\es{cvDef}{
c(z)\equiv a(z)-\frac{z a'(z)}{2(d-1)}\,, \qquad v\equiv {\dot\rho\ov \sqrt{1+{(\pa_\Omega \rho)^2\ov \rho^2}}}\,,
}
with these from \eqref{zEOM}, $z(t,\Omega)$ is given by
\es{zEOM2}{
z(t,\Omega)=c^{-1}\le(v^2(t,\Omega)\ri)\,.
}
We can plug this back into (the scaled version of) \eqref{AreaFunct}, and obtain the membrane theory action
\es{MembraneTheory}{
S[A(T)]=s_\text{th}\min_{m\sim A(T)/(t=0)} \int_m  dt\, d\Omega\ \rho^{d-2}\sqrt{1+{(\pa_\Omega \rho)^2\ov \rho^2}}\, {\mathcal{E}(v)}\,,
}
where we set $\Lambda=1$, introduced $s_\text{th}\equiv{1\ov 4G_N}$, and took into account the fact that since we are studying only a section of the HRT surface, this section ends on the $t=0$ slice, hence we have to use the criterion of relative homology. The function $\mathcal{E}(v)$ encodes the bulk geometry:
\bea\label{E(v)}
\mathcal{E}(v)= \sqrt{ \frac{-a'(z)}{2(d-1)z^{2d-3}}}\,\,\Bigg\vert_{z=c^{-1}(v^2)}\,.
\eea
 The minimization problem concerns the function $\rho(t,\Omega)$ which determines a codimension-1 surface, a membrane in $[0,T]\times \mathbb{R}^{d-1}$ that stretches between the upper and lower ends of the time interval: it is anchored on $\pa A(T)$ on the upper end and ends perpendicularly on the lower end. This membrane is the projection of the HRT surface to the boundary along constant infalling time $t$. Remarkably, if we solve the membrane minimization problem given by the action \eqref{MembraneTheory}, then we can reconstruct the relevant part of the HRT surface  using the map \eqref{zEOM2} (to leading order in the scaling limit). 
  
The function $\mathcal{E}(v)$ can be proven to be a positive even function, monotonically increasing and convex for $0\leq v \leq v_B$, and at $v=v_B$, with $v_B$ the butterfly velocity, it obeys $\mathcal{E}(v_B)=v_B,\, \mathcal{E}'(v_B)=1$ \cite{Mezei:2018jco}. These properties are required for physical consistency \cite{Jonay:2018yei}. A consequence of this is that minimal membranes have $\abs{v}\leq v_B$ for all shapes $A(T)$ \cite{Mezei:2018jco}.\footnote{We have that $z(v_B)=1$, the horizon. Then this fact about minimal membranes translates into HRT surfaces staying behind the horizon in the scaling limit, as we assumed for the derivation.} The membrane tension $\mathcal{E}(v)$ given in \eqref{E(v)} diverges as $v\to 1$. For ease of presentation, we will modify the membrane theory to include a modified membrane tension
\es{tildeE(v)}{
 \widetilde{\mathcal{E}}(v)\equiv\begin{cases}
 \mathcal{E}(v)\,, \quad &(\abs{v}\leq v_B)\,,\\
\abs{v}\,, \quad &(\abs{v}\geq v_B)\,.
\end{cases}
}
The modified membrane problem still has the same minimal membranes as the original ones, because minimal membranes always have $\abs{v}\leq v_B$, in which regime the tensions agree. 

Next we rewrite \eqref{MembraneTheory} in a more geometric manner. First, we want to find a geometric interpretation of $v$. To this end, we write the local co-normal of the membrane $n_\mu$. The membrane is defined by
\es{membraneDef}{
0=\phi(t,\rho,\Omega)\equiv \rho-\hat{\rho}(t,\Omega)\,,
}
where $\hat{\rho}(t,\Omega)$ is just the shape of the membrane in spherical coordinates. The normal covector is then
\es{NormalVec}{
n_\mu=n_\rho \pa_\mu \phi=n_\rho\le(-\dot\rho,1,-\pa_\Omega \rho\ri)\,,
}
where we omitted the hat from $\hat{\rho}(t,\Omega)$. $n_\rho$ is determined by the normalization condition $n^2=1$ to be
\es{nrho}{
n_\rho={1\ov \sqrt{\dot\rho^2+1+{\le(\pa_\Omega \rho\ri)^2\ov \rho^2}}}\,.
}
Note that we are using Euclidean signature, even though the field theory naturally lives in Minkowski signature. Since we have a preferred time in this problem (and no Lorentz invariance), this is merely a choice. Now we can express $v$ in terms of the normal vector and the timelike unit vector $\hat{t}$ in different ways:
\es{vExpr}{
v={n_t\ov \abs{\vec{n}}}=\frac{ (n\cdot \hat{t})}{\sqrt{1- (n\cdot \hat{t})^2}}\,,
}
and rewrite \eqref{MembraneTheory} in a more geometric way: 
\es{MembraneTheory2}{
S[A(T)]&=s_\text{th}\min_{m\sim A(T)/(t=0)} \int_m  d^{d-1}\sigma \ \sqrt{h}\, {\widetilde{\mathcal{E}}(v)\ov \sqrt{1+v^2}}\\
&=s_\text{th}\min_{m\sim A(T)/(t=0)} \int_m  d^{d-1}\sigma\ \sqrt{h}\, \abs{\vec{n}}\,{\widetilde{\mathcal{E}}\le({n_t\ov \abs{\vec{n}}}\ri)}\,,
}
where $\sigma$ are coordinates on $m$ and $h_{\alpha\beta}(\sigma)$ is the induced metric on $m$. The first line was the form of the membrane theory given in the Introduction in \eqref{MembraneTheoryIntro}.

\subsection{Maximin in the scaling limit\label{section-maximin}}
The original derivation of the membrane theory relies on the HRT prescription  \cite{Hubeny:2007xt} for the computation of the holographic entanglement entropy in dynamical settings. The covariant generalization of the bit threads arises naturally by taking the maximin reformulation of the HRT  prescription \cite{Wall:2012uf} as a starting point. Hence,  as a preparation for the discussion of the bit thread analogue of the membrane theory, we present the problem from the perspective of the maximin
prescription.

First let us recall the maximin formula. The entanglement entropy of a boundary region $A(T)$, $S[A(T)]$ is given by
\bea \label{maximin}
S[A(T)]={1\ov 4 G_N}\underset{\Sigma \supset \partial A(T)  }{\textrm{max}}\,\, \underset{\underset{m\subset \Sigma}{m \sim A(T)}}{\textrm{min}} \,\,{\rm area}(m)\,,
\eea 
where the prescription requires us to pick a Cauchy slice $\Sigma$ which contains the boundary $\partial A(T)$ at the boundary of the spacetime $\partial M$, and for each $\Sigma$ find the minimal surface $m\sim A(T)$ contained in it. Then, we vary over all the space-like Cauchy surfaces $\Sigma$ until we reach the maximum among all the minimal surfaces, the maximin surface. Its area over $4G_N$ equals the holographic entanglement entropy. We would like to understand how the above prescription simplifies in the scaling limit.

First, let us recall, that the bulk region of interest to us lies  behind the horizon in the geometry \eqref{f-metric}. In the coordinates used there, this region obeys $t>0$ and $z>z_h=1$. As explained above, for our purposes the boundary region $A(T)$ will lie on the $z=1$ horizon and its boundary $\partial A(T)$ will lie at the intersection $(z=1) \cap \(t=T\)$.

 A Cauchy surface $\Sigma$ in these coordinates can be parametrized by a function $z=\hat{z}(t,\vec{x})$ which can be described as the zero level set of a scalar function $\phi$, given by
\bea
\phi(x^a)\equiv z-\hat{z}(t,\vec{x})
\eea
in terms of the full bulk spacetime coordinates $x^a=(t,\vec{x},z)$. Up to normalization, the normal covector can be obtained from the above scalar function via
\bea\label{normal}
n_a={\cal N} (\partial_a \phi)={\cal N}\le(-\partial_t z,\,  -\partial_i z ,\, 1\ri)\,,
\eea
which we evaluated at the surface $\phi=0$ and hence dropped the hat from $z$. The condition on the hypersurface to be spacelike is simply stated as $g^{ab}\partial_a \phi \partial_b \phi<0$, which gives
\es{Spacelike}{
 \(a(z)-\frac{2 \partial_t z}{b(z)} +\frac{|\nabla z|^2 }{b^2(z)}\) <0\,.
}
Now let us rescale the boundary theory coordinates as $(t,\vec{x})\to \(\Lambda\,t,\,\, \Lambda\,\vec{x} \)$ but leave $z$ unscaled. This choice corresponds to the function $z(t,\vec{x})$ only varying slowly. Then \eqref{Spacelike} becomes
\es{Spacelike2}{
 \(a(z)-\frac{2 \partial_t z}{b(z)\Lambda} +\frac{|\nabla z|^2 }{b^2(z)\Lambda^2}\) <0\,,
}
which is then identically true behind the horizon for any $z(t,\vec{x})$. Let us also write the  metric (\ref{f-metric}) after rescaling:
\bea\label{f-metric-2}
ds^2=\frac{\Lambda^2}{z^{2}}\left( -a(z)dt^{2} +d\vec{x}^{2}-\frac{2}{b(z)\, \Lambda}dt dz  \right)\,,
\eea
which to leading order becomes degenerate, i.e. we can move freely in the $z$ direction.

After these preparations, let us write the maximin procedure after the rescaling.
The space of induced metrics $g_{\mu \nu}(z)$ relevant for the first step of the maximin prescription (the minimization step), is parametrized by a single scalar function $z(t,\vec{x})$. Additionally, we can pick coordinates $x^\mu\in \Sigma_0$ (one can think of $\Sigma_0$ as the horizon, the boundary of $M$ with $z=1$),  which describe points on any hypersurface $\Sigma$ determined by $z(x^\mu)$ in a global way, that is independently of the specific function $z(x^\mu)$.  In these coordinates, one can write down explicitly the maximin formula as
\bea \label{maximin2}
S[A(T)]=\underset{z(T, \partial A(T))=1}{\underset{z(t,\vec{x}) }{\max}}\,\, \underset{ m \sim A(T)}{\rm{min}} \,\,\, \frac{1}{4 G_N}\int_{m} d^{d-1}x_m
\ \sqrt{h(z(x_m),x_m)}\, \eea 
where the surface $m$ is described simply by a set of points $x_m^\mu$, in $\Sigma_0$.  

The two optimization steps in the above program are independent of each other, and therefore we can exchange their order, and equivalently compute $S[A(T)]$ as 
\bea \label{maximin3}
S[A(T)]= \underset{ m \sim A(T)}{\rm{min}} \,\, \underset{z(m)}{\textrm{max}}\,\,\, \frac{1}{4 G_N}\int_{m}d^{d-1}x_m\ \sqrt{h(z(x_m),x_m)}\,  \,,
\eea 
where now the maximization step is carried out on the restriction of $z(x)$ on $m$, this is $z(m)$. Taking \eqref{maximin3} as our starting point, one can explicitly solve the equations of motion for $z(m)$ as done in  \cite{Mezei:2018jco} and reviewed in Section~\ref{section2}, which gives us the solution to the maximization step above. Plugging the solution back into (\ref{maximin3}) gives us the membrane theory description \eqref{MembraneTheory2}. In Appendix~\ref{app:maximin} we provide an alternative viewpoint on the equivalence of the programs \eqref{maximin2} and \eqref{maximin3}.

Away from the hydrodynamic limit, we cannot exchange the optimization steps in maximin. We found it worth listing some obstacles that prevent us from switching the order in the general case and how they get resolved in the limit we are studying:
\begin{itemize}
\item The induced metric on a codimension one surface $\(m, z(m)\)$ lying on the hypersurface $z(x)$ depends not only on $z(x)$ but also on its derivatives $\partial_\mu z(x)$ evaluated at $m$. This is not the case in (\ref{maximin2}) and (\ref{maximin3}), since contributions to the induced metric proportional to $\partial_\mu z$ are suppressed in the scaling limit, as can be seen from (\ref{f-metric-2}).  

\item In \eqref{maximin3}, $z(m)$ is an arbitrary function only constrained by the boundary condition at $\partial A(T)$. One may worry about the existence of a spacelike hypersurface $z(x)$ that matches it on $m$. The scaling limit comes to the rescue: as argued around (\ref{Spacelike2}) any surface $z(x)$ is spacelike  and therefore an arbitrary surface $z(m)$ can always be embedded in a spacelike hypersurface $z(x)$.

\item The maximization step in (\ref{maximin3}) can be interpreted as the area maximization of a codimension one hypersurface $z(x_m)$ in the manifold $m\times [1,\infty)$. Had this manifold been Riemannian, a maximal area hypersurface would not exist, since one could always increase the area by wiggling the surface along the embedding direction. From (\ref{f-metric-2}) we can read off that the metric on $m\times [1,\infty)$ is degenerate, hence not Riemannian, and the  arbitrary increase in area from wiggling is avoided in the scaling limit due to the fact that the induced metric on $m$ becomes independent of $\partial_\mu z$.

\end{itemize}

\section{Bit-threads in the scaling limit \label{2.2}}

\subsection{Bit-threads for the membrane theory \label{3.2}}

Having reviewed the derivation of the membrane theory, next we are interested in deriving a dual concave max flow program whose solution equals the solution of \eqref{MembraneTheory2}. There is a powerful technique that allows to connect solutions of convex mincut-like programs similar to the minimization problem described above with solutions of concave max-flow programs called convex dualization. However, the min cut-like problem \eqref{MembraneTheory2} is not in any obvious way a convex optimization problem. Nevertheless, in some circumstances as the ones studied in \cite{Headrick:2017ucz,Harper:2018sdd,Cui:2018dyq}, it is possible to recast the solution of a non-convex problem in terms of a convex program by a trick called {\it  convex relaxation}.  Convex relaxation is a technique in which we embed a non-convex program into a convex one in such a way that the solution of the latter equals that of the former. For a recent and concise review of all the relevant concepts from convex optimization used in this section, we recommend  \cite{Headrick:2017ucz}. Our presentation will follow closely their notation and derivations adapted to our problem of interest. 

First, we want to rewrite the membrane minimization into a minimization over the whole manifold by ``smearing'' the membrane.
To this end, consider a once differentiable map (the smearing) $\psi: M\to \mathbb{R}$ with the boundary condition $\psi|_{\partial M}=\chi_{A}$, where $\chi_A$ is $1$ on $A$ and $0$ on $A^c$. (One can make $\psi|_{\partial M}$ also differentiable by approximating the step like function $\chi_A$ arbitrarily well with smooth functions.)  This implies that the preimage $\psi^{-1}(p)$ with $p \in [0,1]$ is a codimension one surface homologous to $A$, see Section 3.2 of \cite{Headrick:2017ucz} for a more detailed analysis. The level set functions given by the preimage of $\psi$, $\psi^{-1}(p)$ for arbitrary $p$ have unit conormal given by 
\bea
n_\mu=\frac{\partial_\mu \psi}{|\partial \psi|}\,.
\eea
We would like to show that 
\bea\label{min-cut}
\underset{\psi\,, \psi|_{\partial M}=\chi_A}{\rm min} \int_M  \mathcal{L}(n) |\partial \psi | =  \underset{ m \sim A}{\rm{min}}  \int_{m} \sqrt{h_{m}} \,\,  \mathcal{L}(n) \,,
\eea 
where we introduced
\bea
\mathcal{L}(n)\equiv |\vec{n}|\,\widetilde{\mathcal{E}}\( \frac{n_t}{|\vec{n}|} \)\,.
\eea
In (\ref{min-cut}), $h_m$ represents the determinant of the induced metric on the surface $m$.

\begin{center}
\begin{figure}[!h]
\centering
\includegraphics[scale=0.6]{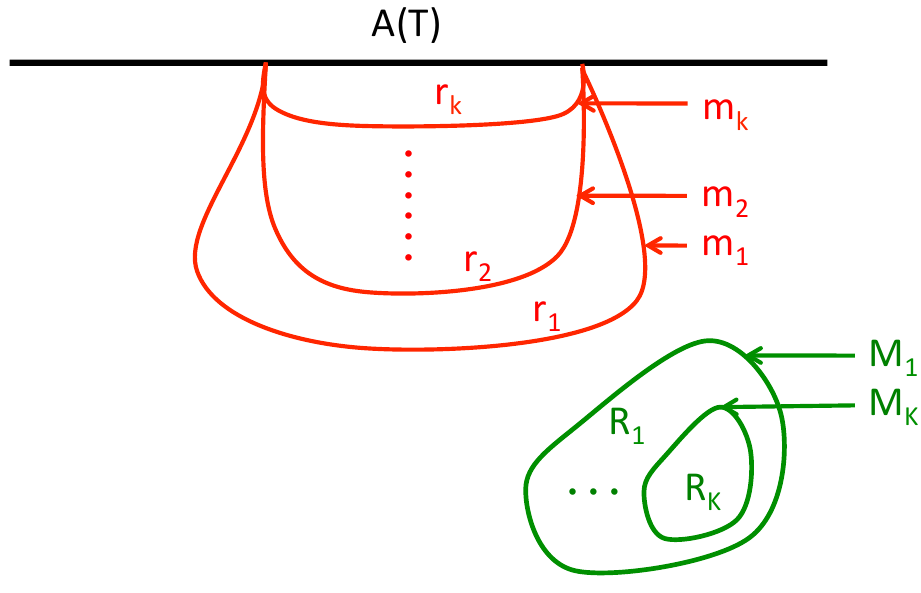}
\caption{The regions $r_i\sim A$, while $R_I\sim \emptyset $. The surfaces are the boundaries of the respective regions, $m_i=\pa r_i \setminus A(T)$ and $M_I=\pa R_I$. The condition $\sum_{i=1}^k \lambda_i=1$ comes from the requirement that $\psi\vert_{\pa M}=\chi_A$.
\label{fig:StepFunct}}
\end{figure}
\end{center}

The argument is simple. Let us approximate $\psi$ with a sum of step functions such that it also satisfies the boundary conditions:
\es{PsiStep}{
\psi(x)&=\sum_{i=1}^k \lambda_i \, \chi_{r_i}(x)+\sum_{I=1}^K \zeta_I\,  \chi_{R_I}(x)\,,\\
\sum_{i=1}^k \lambda_i&=1\,;
}
see Fig.~\ref{fig:StepFunct} for an explanation of what these regions mean. Let us also introduce $E[m]\equiv \int_{m} \sqrt{h_{m}} \,\,  \mathcal{L}(n)$.   Since $\pa \psi$ is a sum of delta functions, the LHS of \eqref{min-cut} becomes:
\es{MinCutLHS}{
\min_{\underset{\zeta_I,\, M_I}{\lambda_i,\, m_i}}\le(\sum_{i=1}^k  \abs{\lambda_i} E[m_i]+\sum_{I=1}^K \abs{\zeta_I} E[M_I]\ri)\,.
}
Since $E[m]\geq0$, minimization in $\lambda_i,\, \zeta_I$ for fixed $m_i,\, M_I$ gives $\zeta_I=0,\, \lambda_{i^*}=1,\,\lambda_{i\neq i^*}=0$ where $E[m_{i^*}]$ is the minimal among the $E[m_i]$. Then \eqref{MinCutLHS} becomes $\min_m E[m]$, which is exactly what we have on the RHS. We note that our pedestrian proof can be made more concise and rigorous following the route taken in \cite{Headrick:2017ucz}; we preferred the version presented here for pedagogical reasons.

The above analysis show that one can write the entanglement entropy in terms of the following minimization problem
\bea\label{SA-2}
S[A(T)]=s_{th} \,\underset{N}{\rm min}\,  \int_M \,\mathcal{L}(\hN)\,|N| \,,
\eea
where $\hN\equiv {N\ov \abs{N}}$, and the minimization problem is subject to $N=\partial \psi$ with $\psi$ obeying the boundary condition $\psi|_{\partial M }=\chi_A$.

Second, we realize that \eqref{SA-2} is a convex optimization program: the equality constraints and the objective function are convex. To show the convexity of the latter, we compute the eigenvalues of its Hessian matrix $\pa_i\pa_j \le(\mathcal{L}(\hN)\,|N| \ri)$:
\es{HessianEigenvalues}{
\le\{0\,, \quad {\widetilde{\mathcal{E}}\le(v\ri)- v\, \widetilde{\mathcal{E}}'\le(v\ri)\ov\abs{\vec{N}} } \, \quad [\text{$(d-2)$-fold degenerate}]\,, \quad {\le(1+ v^2\ri)\, \widetilde{\mathcal{E}}''\le(v\ri)\ov\abs{\vec{N}} }\ri\}\,,
}
where we used the notation $v={N_t/ \abs{\vec{N}}}$ to avoid clutter. By the properties of $ \widetilde{\mathcal{E}}\le(v\ri)$ that we listed above \eqref{tildeE(v)} these eigenvalues are positive semidefinite, thus we showed that the objective function is convex.

The next step is to dualize the program in order to derive the max flow analogue of (\ref{SA-2}), and then argue that both programs share the same solution. The precise arguments behind the equality between the solution of a convex program and the solution of its dual concave program were beautifully presented in \cite{Headrick:2017ucz} and applied explicitly to the RT formula. We will rely heavily on the terminology and presentation of \cite{Headrick:2017ucz}, and simply quote the main theorems and results as the derivation progresses.

The procedure known as {\it convex dualization} starts with a given convex (concave) program, defined by an  {\it objective function}, which is the function we want to minimize (maximize), and a set of inequality and equality constraints imposed on the domain on which the objective function is defined, and defines the set, on which one wants to find the optimal solution, {\it the feasible set}. Then, one constructs a {\it Lagrangian function}, which is obtained from the objective function by adding {\it Lagrange multipliers} to impose the equality and inequality constraints explicitly.  In many situations this allows the optimization of the original objective function, and as a result, it reduces the Lagrangian function to a function of the Lagrange multipliers, which then can be interpreted as a concave (convex) function on its own.   The resulting program that this function defines is called the concave (convex) {\it dual program}. See  \cite{Headrick:2017ucz} for a detailed explanation of this procedure as well as some relevant examples. 

 As a consistency condition for the convex dualization to work it is required that at every step the Lagrangian function should have finite minimum (maximum) with respect to the original variables and finite maximum (minimum) with respect to the Lagrange multipliers. This consistency condition defines the set of values that the dual function can have and therefore defines the {\it dual feasible set}. A sufficient condition ({\it Slater's condition}) for the solution of the original {\it primal } program  to equal the solution of the dual program is the existence of an element in the feasible set for which all the inequality constraints become strict inequalities. We will apply this procedure to equation (\ref{SA-2}). In the remainder of this section we will assume that the reader is familiar with the concepts used in the previous paragraphs at the level and in the context of \cite{Headrick:2017ucz} .

We will ignore the overall coefficients in  (\ref{SA-2}) and write the relevant {\it primal} program as 
\es{primal}{
{\rm minimize} \quad  \int_M \,\mathcal{L}(\hat{N})\,|N|, \quad {\rm subject \,\,\,\,  to }\quad N_\mu=\partial_\mu \psi \, \quad {\rm and } \quad   \psi|_{\partial M}=\chi_A\, .
}
The Lagrangian function associated to it is 
\bea\label{Lagrangian}
L[N,\psi; w , \phi]= \int_M \mathcal{L}(\hat{N})|N|+\int_M w^\mu(N_\mu -\partial_\mu \psi)+\int_{\partial M}\sqrt{h}\phi\(\psi-\chi_A\)\,,
\eea
where we  added the Lagrange multipliers $w^\mu$ and $\phi$ to enforce explicitly the equality constraints. Maximizing over them gives back the primal program \eqref{primal}. Instead we will exchange the order of maximization in $w^\mu,\phi$  and minimization in $N, \psi$.

Minimizing the above functional with respect to the original variables $N, \psi$  give rise to a dual objective function $G[w,\phi]$ which depends on the Lagrange multipliers $w$, $\phi$ and defines the concave dual program. In order to study this function, let us rewrite the Lagrangian function (\ref{Lagrangian}) by moving the partial derivative on $\psi$, and collecting the boundary terms proportional to $\psi$. We consider its minimum with respect to the primal program variables 
\es{minLag}{
G[w,\phi]&=\,\underset{\psi, N}{\rm min}  \Big[ \int_M \( \mathcal{L}(\hat{N})+w^\mu \hat{N}_\mu \)|N|+ \int_M  \psi \(\nabla_\mu w^\mu\)  \\
& \qquad \qquad + \int_{\partial M}\,\sqrt{h}\, \psi\, \( \phi +n_{\partial M}^\mu w_\mu\) -\int_{\partial M}\sqrt{h}\,\phi \,\chi_A  \Big]\,.
}
We need the minimum of this function, with respect to $\psi$ and $N$ to be finite so that $w^\mu$ and $\phi$ are in the domain of $G[w,\phi]$. This requirement imposes the following constraints on the Lagrange multipliers: 

{\bf First:} The term in parenthesis in the first integral of (\ref{minLag}) cannot be negative. Otherwise, by chosing $|N|$ arbitrary large one could have arbitrary large negative contributions from it,  and  the function $G[w, \phi]$ then will not be finite. This means that $w$ must be such that 
\bea\label{Nw}
\mathcal{L}(\hat{N})+w^\mu \hat{N}_\mu \geq 0
\eea
for all values of $\hat{N}_\mu$. This imposes a non-trivial constraint on $w$ which we will study later.

{\bf Second: }The second integral of (\ref{minLag}) imposes the  condition 
\bea\label{div}
\nabla_\mu  w^\mu=0\,:
\eea
the Lagrange multiplier field $w^\mu$ must be divergenceless. Otherwise, by choosing $\psi$ such that $|\psi|$ is large and opposite in sign to $\nabla_\mu  w^\mu$, we could have arbitrary large negative contributions from this term.

{\bf Third:} In complete analogy with the second integral, the third integral, which is defined on the boundary $\partial M$, will allow arbitrary large negative contributions unless 
\bea\label{phi}
\phi=-n_{\partial M}^\mu w_\mu\,.
\eea

Imposing constraints (\ref{Nw}), (\ref{div}), and  (\ref{phi}) implicitly in (\ref{minLag}), leads to the following dual objective function
\bea \label{dual1}
 \int_A\,\sqrt{h}\, w^\mu n_\mu\,,
\eea
where $w$ is a divergenceless vector field, which satisfies the condition
\bea\label{Nw-2}
\mathcal{L}(\hat{N})+w^\mu \hat{N}_\mu \geq 0
\eea
for any vector field $\hat N_\mu$ with unit norm. The dual program is the maximization of the dual objective function on the feasible set defined by the above constraints. The dual program is therefore a concave max flow program as the primal program was a convex min cut-like program. Next, we will describe this program in a more refined way.

The dual objective function (\ref{dual1}) is obtained by carrying out the minimization step of (\ref{minLag}) in the  dual feasible set.  That is, condition  (\ref{Nw}) implies that the first integral is positive definite. Therefore, one can achieve its minimum by setting $|N|=0$ everywhere in the bulk. Note that this leaves $\hat{N}$ (the direction of $N$) undetermined. Likewise, the second and third integrals are identically zero by the dual constraints (\ref{div}) and (\ref{phi}). The only remaining term is  the last boundary integral which on the feasible set equals (\ref{dual1}). 

$w$ is in the allowed set, if it satisfies \eqref{Nw-2} for any unit norm one-form $\hat{N}$. To find this set, we minimize
the LHS of (\ref{Nw-2}) with respect to $\hat{N}_\mu$. The optimal $\hN$, $\hN^*$ would depend on $w$ and so we refer to it as $\hN^*(w)$. Plugging the optimal $\hN^*(w)$ back into (\ref{Nw-2}) defines for us the set of allowed $w$'s. Notice that finding the optimal $\hN^*(w)$ is equivalent to choosing the $\hat{N}$ that takes the inequality as close as possible to saturation. Indeed, it should be always possible to find an optimal $w^*$ for which the inequality is saturated, that is 
\es{}{
\mathcal{L}(\hat{N}^*(w^*))+(w^*)^\mu \hat{N}^*_\mu (w^*)= 0\,,
}
such $w$ defines the boundary of the feasible set.

Now, the direction of the full covector $N_\mu$ can be described by the following data:
\bea
v=\frac{N_t}{|\vec{N}|}, \quad {\rm and} \quad n_i=\frac{N_i}{|\vec{N}|}
\eea
where $n_i$ satisfies the constraint $|\vec{n}|=1$, then, the full vector have two different presentations in terms of these variables, these are:
\bea\label{Nmu-5}
N_\mu=|N|\(\frac{v}{\sqrt{1+v^2}}\,, \frac{n_i}{\sqrt{1+v^2}}\) \quad {\rm or} \quad N_\mu=|\vec{N}|\(v, n_i \)\,.
\eea
Then, one can vary $N_\mu$ with respect to $(v,n_i)$, while keeping either $|N|$ or $|\vec{N}|$ fixed, using the first or the second presentation of $N_\mu$ in (\ref{Nmu-5}) respectively.  In the minimization of the first term of (\ref{minLag}) this is equivalent to either varying the term 
\bea\label{presentations-1}
\( \mathcal{L}(\hat{N})+w^\mu \hat{N}_\mu \) \quad {\rm or }\quad \(\widetilde{\mathcal{E}}(v)+w^t v +w^i\,n_i\)
\eea
with respect to $(v,n_i)$, since 
\es{presentations}{
 \mathcal{L}(N)+w^\mu N_\mu &=|N|\( \mathcal{L}(\hat{N})+w^\mu \hat{N}_\mu \)\, \\
 &=|\vec{N}|\(\widetilde{\mathcal{E}}(v)+w^t v +w^i\,n_i\) 
}
 and
 \bea
|N|=\sqrt{1+v^2}\,|\vec{N}|.
\eea 
Then, the constraint (\ref{Nw-2}) is equivalent to
\bea\label{Nw-3}
\(\widetilde{\mathcal{E}}(v)+w^t v +w^i\,n_i\)
\geq 0\,.
\eea
We chose to minimize the LHS of the above expression instead, since its minimum is straightforward to obtain\footnote{Notice that $|N|\to 0$ when $|\vec{N}|\to 0$ for any $\(v, n_i\)$ and they are both positive definite. Therefore, in the minimization of (\ref{presentations}) over $v$ we can either keep $|N|$ or $|\vec{N}|$ fixed without affecting the region of allowed $w$. }.
As a function of $n_i$ the minimum is achieved with 
\bea\label{op-ni}
n^*_i=-\frac{w_i}{|\vec{w}|}
\eea
and as a function of $v$ it is achieved with 
\bea\label{op-v}
-\widetilde{\mathcal{E}}'(v^*)=w^t.
\eea
In the convex domain of $\widetilde{\mathcal{E}}(v)$ which is for $0 \leq |v| \leq v_B$, the function $\widetilde{\mathcal{E}}'(v)$ is one to one, and therefore invertible, this allows us to think of $v^*$ as $v^*=v^*(w^t)$ in (\ref{op-v}) by formally inverting the equation. (For $\abs{v}>v_B$ we have that $\widetilde{\mathcal{E}}'(v)=1$, which is not invertible.) This restricts the range of $w^t$ to $-1\leq  w^t \leq 1$.

Plugging (\ref{op-ni}) and (\ref{op-v}) back into the inequality constraint (\ref{Nw-3}), one gets
\bea
\(\widetilde{\mathcal{E}}(v^*)+w^t v^* +w^i\,n^*_i\)=\( \widetilde{\mathcal{E}}(v^*)+ v^*\, w^t -|\vec{w}|\)\geq 0\,.
\eea
The above inequality leads to the following constraint on $w$
\bea\label{norm-constraint-2}
|\vec{w}|\leq  H(w^t)\,,
\eea
where $H(w^t)$ is the Legendre transform of $-{\cal E}(v)$:\footnote{The same function has appeared in \cite{Jonay:2018yei}, where in the one-dimensional case it was interpreted as an entropy production rate, and it was denoted by $\Gamma\le(\pa S\ov \pa x\ri)$. It would be interesting to relate the two pictures.}
\es{Legendre-0}{
w^t(v)&=-\mathcal{E}'(v)\,,\\
H(w^t)&=\mathcal{E}(v)-v\,\mathcal{E}'(v)\Big|_{v=v(w^t)}\,.
}
We have removed the $*$ super index in $v$ since at this stage $v$ is simply a convenient parametrization of the norm bound constrain of $w$.
 
In order to have a concave program with a well defined global maximum we need the constraints of the program to be concave. The divergenceless constraint of $w$ is an affine constraint and therefore trivially concave. Similarly, the norm constraint (\ref{norm-constraint-2}), when written as 
\bea\label{norm-constraint-3}
H(w^t)-|\vw| \geq 0\,,
\eea
is clearly concave provided the functions $H(w^t)$ and $-|\vw|$ are concave with respect to their arguments. As mentioned earlier, $H(w^t)$ is the Legendre transform of $-\mathcal{E}(v)$, and $-\mathcal{E}(v)$ is a concave function, then, since Legendre transformation takes concave functions to concave functions, we conclude that $H(w^t)$ is concave as well, while concavity of $-|\vw|$ follows from the triangle inequality. Therefore, the norm constraint (\ref{norm-constraint-3}) is a concave constraint. 

\begin{figure}
\centering
\includegraphics[scale=0.7]{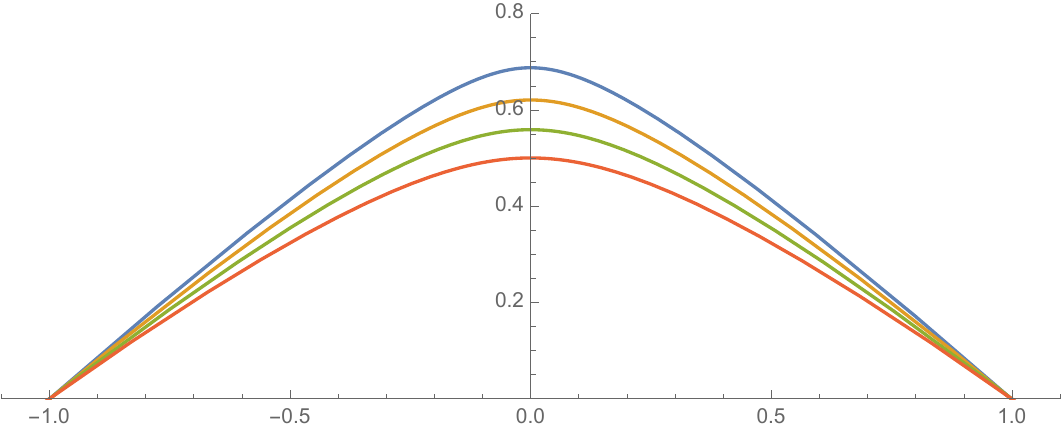}
  \setlength{\unitlength}{1cm}
  \setlength{\unitlength}{1cm}
\begin{picture}(0,0)
\put(-7.4,5.4){$H(w^t(v))$}
\put(-0.35,0){$w^t(v)$}
\end{picture}
 \caption{Upper bound on $|\vec{w}|$ as a function of $w^t$. The plot is made parametrically, using $v$ as the parameter. The function $H(w^t(v))$ attains its maximum value at $v=w^t=0$, where it equals $v_E$, which is monotonically decreasing as a function of $d$ and so is the function $H(w^t(v))$. We plot the curves for $d=3,4,7,\infty$.  \label{vbound}}
\end{figure}
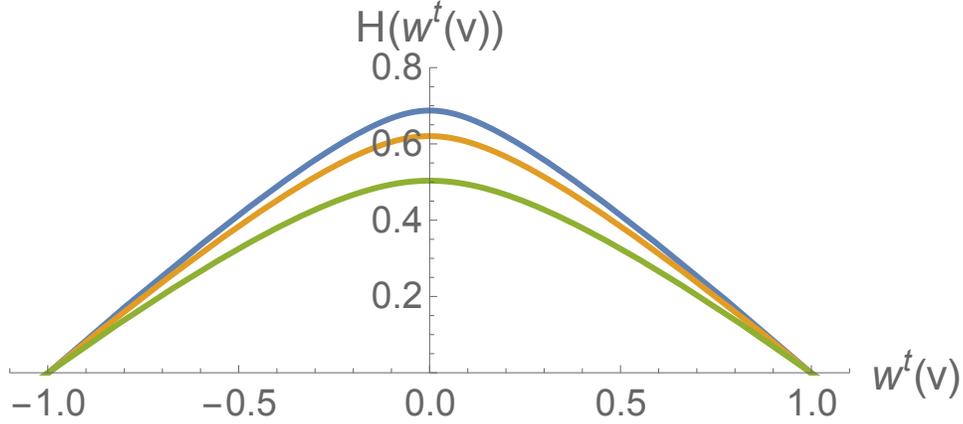

In order to gain some intuition about the norm bound (\ref{norm-constraint-2}), we consider the explicit case of black holes with zero chemical potential \cite{Mezei:2018jco}, and illustrate the norm bound (\ref{norm-constraint-2}) in Fig~\ref{vbound}. In this case 
\es{}{
\mathcal{E}( v)=\frac{v_E}{(1-v^2)^{(d-2)/(2d)}},
\quad{\textrm{where}}\quad v_E=\frac{\(\frac{d-2}{d}\)^{(d-2)/(2d)}}{\(\frac{2(d-1)}{d}\)^{(d-1)/d}}\quad \textrm{and} \quad v_B=\sqrt{\frac{d}{2(d-1)}}\,. 
}
The boundary of the convex domain of $w$ is given parametrically by the functions 
\es{}{
H(w^t(v))&=\mathcal{E}( v)-v\mathcal{E}'( v)=\frac{v_E}{\(1-v^2\)^{\frac{3d-2}{2d}}} \(1-\frac{v^2}{v^2_B}\)\,,  \\
 w^t(v)&=-\mathcal{E}'( v)=-\frac{d-2}{d}\,\frac{v_E\, v}{\(1-v^2\)^{\frac{3d-2}{2d}}}\,,
}
from which the expected behavior can be read off. That is, $H(w^t(v))$ goes to zero as $v\to \pm v_B$ and equals $v_E$ at $v=0$, while $w^t(v)$ varies monotonically as a function $v$.

In summary, we  found that the max flow program dual to the minimal ``energy'' of the membrane theory is given by the following flow maximization:
\es{flux-membrane}{
S[A(T)]=s_{th} \,\underset{w}{\rm max}\int_{A(T)}\,\sqrt{h}\, w^\mu n_\mu\,,\quad \text{subject to}\quad \nabla_\mu w^\mu=0\,,\quad {\rm and } \quad   |\vec{w}|\leq H(w^t)\, ,  
}
where $H(w^t)$ is the Legendre transform of $-{\cal E}(v)$:
\es{Legendre}{
w^t(v)&=-\mathcal{E}'(v)\,,\\
H(w^t)&=\mathcal{E}(v)-v\,\mathcal{E}'(v)\Big|_{v=v(w^t)}\,. 
}
The membrane program is clearly concave, as the objective function is linear in $w^\mu$ and so is the divergenceless constraint, while the norm constraint is concave. Thus, we obtained a well defined convex optimization program.

Finally, the max flow program just described admits the field configuration $w^\mu=0$ as an element of the feasible set, since it satisfies both the divergenceless condition (trivially), and importantly the  inequality constraint associated to the norm bound strictly, i.e. $H(w^t)-\abs{\vec{w}}\vert_{w=0}>0$. This means that the dual program obeys {\it Slater's condition},  and therefore its solution equals that of the primal program.

 \subsection{Max flow - min cut theorem and the meaning of $v$ in the bit thread problem} \label{sec:vMeaning}
In the formulas \eqref{Legendre}, $v$ appears simply as a convenient parameter to describe the norm bound on $w$ and therefore in the bit thread language it does not  have an obvious geometric interpretation on its own. It will prove useful  to uncover  a geometric interpretation by studying the max flow program in more detail. 

First, let us notice that the obvious generalization of the max flow - min cut theorem  of  \cite{Freedman:2016zud} in this context tell us that the bound on $w$ is saturated precisely at the co-dimension one surface $m(A(T))$ at which the min cut program (\ref{MembraneTheory2}) achieves its minimum (minimum energy of the membrane), and it does it in such a way that the flux integrand is as large as possible. We will lay out the usual arguments leading to this expectation.

Starting from the max flow program (\ref{flux-membrane}) one can write down a global upper bound on the integral by maximizing its integrand pointwise. That is 
\bea \label{mem-MFMC-1}
 \underset{w}{\rm max}\int_m\,\sqrt{h}\, w^\mu n_\mu \leq  \,\, \underset{w_*}{\rm max} \int_m\,\sqrt{h}\, w_*^\mu n_\mu\,.
\eea
The maximizations on both sides of the above inequality are taken over vector fields $w$ and $w_*$ subject to the norm bound  (\ref{norm-constraint-3}). However, while $w$ is further restricted to be divergenceless $w_*$ is not. This implies that  the left hand side of (\ref{mem-MFMC-1}) can be evaluated on any surface $m \sim A(T)$, without changing its value, while the right hand side is sensitive to the choice of $m$. Since we want to make the upper bound as tight as possible,  we can further minimize the right hand side over $m$. This results in 
\bea\label{mem-MFMC-2}
 \underset{w}{\rm max}\int_m\,\sqrt{h}\, w^\mu n_\mu \leq  \,\, \underset{m\sim A(T)}{\rm min}  \,\, \underset{w_*}{\rm max} \int_m\,\sqrt{h}\, w_*^\mu n_\mu\,.
\eea
In the rest of the section we prove that the RHS of (\ref{mem-MFMC-2}) is equal to the LHS of (\ref{mem-MFMC})  (once the maximization step is carried out) by analogy with the result of \cite{Freedman:2016zud}. The saturation of the above inequality is equivalent to the equality between the solutions of the primal and dual programs, (\ref{mem-MFMC}). 

Let us consider the maximization problem
\bea\label{flux-membrane-2}
\underset{w_*}{\rm max} \int_m\,\sqrt{h}\, w_*^\mu n_\mu\,,
\eea
where  $w_{*}$ satisfies the norm bound constraint \eqref{norm-constraint-3}. 
Using the notation introduced at the beginning of Section~\ref{2.2}, the unit covector normal to a surface $m$ (not necessarily the minimal ``energy'' membrane) at a given point can be written in terms of the parameter $v\equiv n_t/|\vec{n}|$, in terms of which
\es{nPara}{
n_t=\frac{v}{\sqrt{1+ v^2}} \,,\quad \quad |\vec{n}|=\frac{1}{\sqrt{1+ v^2}}\,.
}
In order to maximize (\ref{flux-membrane-2}) we need the local flux to be as big as possible. There is an easy upper bound on the local flux
 \bea\label{local-flux-term}
 w_{*}^\mu n_\mu=w_*^t n_t+\vec{n}\cdot \vec{w}_*\leq |n_t| |w_*^t|+|\vec{n}||\vec{w}_*|\,,
 \eea
where we have written $w_*=(w_*^t, \vec{w}_*)$.
We can always saturate this bound since there is no constraint on the sign of $w_*^t$ or the spatial direction of $\vec{w}_*$, it is achieved by taking $\vec{w}_*$ to point in the same direction as $\vec{n}$, and sgn$(w_*^t)=$ sgn$(n_t)$. 

We introduce a convenient parametrization of $w_{*}$:\footnote{Note the sign difference compared to \eqref{Legendre}. This is not a typo, when we were minimizing \eqref{presentations} we had to choose $w$ to point in the ``opposite direction'' to $N$, here we are maximizing the flow, hence we choose $w$ to point in the ``same direction'' as $n$. Technically, we are using that ${\cal E}(v)$ is an even function.}
\es{wsParam}{
w_*^t=\lambda\,\mathcal{E}'( \bar{v})\,, \qquad |\vec{w}_*|=\lambda\le(\mathcal{E}( \bar{v})-\bar{v}\mathcal{E}'( \bar{v})\ri)\,,
}
where to satisfy the constraint \eqref{norm-constraint-3}, we have $0\leq \lambda\leq 1$, and $\bar{v}$ is an auxiliary parameter from the range $-v_B\leq \bar{v} \leq v_B$. (One can also gain intuition about this parametrization from Fig.~\ref{fig:intersection}, where changing $\lambda$ corresponds to moving along rays inside the allowed region colored blue.) The virtue of this parametrization is that it covers the allowed set of $w_*$  and that the ratio
\es{wsAng}{
{w_*^t\ov |\vec{w}_*|}=\frac{\mathcal{E}'(\bar{v})}{\mathcal{E}( \bar{v})-\bar{v}\mathcal{E}'( \bar{v})}\equiv {\cal F}(\bar{v})
}
is independent of $\lambda$. This parametrization will be useful throughout the paper. Note that at this stage $v$ and $\bar{v}$ are independent parameters describing the directions of two different vectors, namely $n$ and $w_{*}$. 

With the parametrizations \eqref{nPara} and \eqref{wsParam} we get:
\es{wsn}{
w_{*}^\mu n_\mu &=\(\frac{ w_*^t v+|\vec{w}_*|}{\sqrt{1+ v^2}}\)\\
&=\lambda\(\frac{\mathcal{E}( \bar{v})-(\bar{v}-v)\mathcal{E}'( \bar{v})}{\sqrt{1+ v^2}}\)\,,
}
where we used that $w_*(\bar{v})$ saturates the local flux bound (\ref{local-flux-term}). It is clear that if we want to maximize \eqref{wsn} while respecting (\ref{norm-constraint-3}), we have to set $\lambda=1$, which corresponds to also saturating (\ref{norm-constraint-3}). Finally, we maximize over $\bar{v}$
 (by taking the $\bar{v}$-derivative of \eqref{wsn}), and we find that the maximum is achieved at $\bar{v}=v$, where the flux term is given by 
\bea
 w_{*}^\mu  n_\mu=\frac{ \mathcal{E}( v)}{\sqrt{1+ v^2}}\,,
\eea
and therefore,
\es{wStar}{
\underset{w_*}{\rm max} \int_m\,\sqrt{h}\, w_*^\mu n_\mu\,=\int_m \sqrt{h}\,\,  \frac{ \mathcal{E}( v)}{\sqrt{1+ v^2}}\,.
}
The RHS is a familiar expression: it is  the energy of the minimal membrane in the membrane theory formulation, hence by the results of Section~\ref{3.2}, we have
\es{mem-MFMC}{
 \underset{m\sim A(T)}{\rm min}  \int_m \sqrt{h}\,\,  \frac{ \mathcal{E}( v)}{\sqrt{1+ v^2}}&=
 \underset{w}{\rm max}\int_{A(T)}\,\sqrt{h}\, w^\mu n_\mu\\
 &=\underset{w}{\rm max}\int_{m\sim A(T)}\,\sqrt{h}\, w^\mu n_\mu\,,
}
where  the first line is the statement of duality with the minimal membrane problem (the main result of Section~\ref{3.2}), while in the second line we used that $w$ is divergenceless. Combining \eqref{mem-MFMC} with \eqref{wStar}, we see that in \eqref{mem-MFMC-2} the RHS$=$LHS. This is what we set out to prove.

From the proof it also follows that for the optimal vector fields $w=w_*$ at the minimal surface. Above we determined $w_*$ in terms of the data of the minimal surface $m(A(T))$.  $m(A(T))$ has the normal covector field $n$ living on it, in terms of which
\es{wonm}{
w^t\vert_{m(A(T))}&=\mathcal{E}'( v)\,,\\
\vec{w}\vert_{m(A(T))}  &=\le(\mathcal{E}(v)-v\mathcal{E}'( v)\ri)\sqrt{1+v^2}\,\, \vec{n}\,,
}
with $v={n_t\ov \abs{\vec{n}}}$. We learn that unlike in the case of the minimal surface problem, $w$ is not parallel to $n$, since 
\es{wangle}{
{w^t\ov \abs{\vec{w}}}&={\cal F}(v)\\
&\neq v={n_t\ov \abs{\vec{n}}}\,,
}
where we used the notation introduced in \eqref{wsAng}.
Away from $m(A(T))$ the optimal $w$ is highly nonunique, which corresponds to the flexibility of the bit thread construction of \cite{Freedman:2016zud}.

\section{Relation to bulk bit threads}\label{sec:bulk}

\subsection{From membrane bit threads to bulk bit threads}
In the section (\ref{3.2}) we succeeded in obtaining the flow version of the minimal energy problem appearing in the membrane theory that describes the large scale dynamics of the holographic entanglement entropy.
It is  interesting to ask how the bit threads of (\ref{flux-membrane}) (which we refer to as membrane bit threads) relate to the bulk bit threads which are dual to the HRT picture of entanglement dynamics. One way to answer that question is to find a map between the membrane bit threads studied in this paper and the bulk bit threads familiar from \cite{Freedman:2016zud}. 

First, note that the membrane description of the minimal surface involves a projection to the boundary in which the knowledge about the bulk coordinate $z$, and the curved geometry of the bulk spacetime behind the horizon are encoded in $v$ and $\mathcal{E}(v)$. The explicit maps are derived in appendix \ref{B}, in equations  (\ref{z(v)})  and  (\ref{a(z)}). We recall those here:
\bea\label{zu-1}
z(v)=\(\frac{v}{\mathcal{E}(v) \mathcal{E}'(v)}\)^{\frac{1}{2(d-1)}}\,,
\eea
and 
\bea\label{az-1}
a(z(v))=-v\(\frac{\mathcal{E}(v) -v\mathcal{E}'(v) }{\mathcal{E}'(v) }\)\,.
\eea
This means that in the minimal energy picture of the membrane theory, each surface $m\sim A$ can be mapped to the bulk via (\ref{zu-1}) and (\ref{az-1}), where $v$ is determined from the normal vector on $m$ according to (\ref{vExpr}). 

Second, the optimal membrane bit threads $w_m$ at the bootleneck surface $m$ with local conormal $n$ are fully determined by $n$. We would like to extend this relation for arbitrary membrane bit threads $w_m$, so that we can use this relation to lift the membrane bit threads $w_m$ to bulk bit threads $w_b$ via (\ref{zu-1}). {\it A priori} there are many options how to define $v$ from $w_m$, but from the presentation in Section~\ref{sec:vMeaning} it should be plausible that the correct one is to extend \eqref{wangle} (or \eqref{wsAng}) away from the bottleneck surface $m(A(T))$:
\es{wangle2}{
{w^t\ov |\vec{w}|}&= {\cal F}(v)\,,\quad \implies \quad v={\cal F}^{-1}\le({w^t\ov \abs{\vec{w}}}\ri)\,.
}

To establish the map between membrane bit threads and bulk bit threads, let us consider a bulk hypersurface parametrized by the function $z(x)$ as was done in Section~\ref{sec:membrane}, where the intrinsic coordinates $x$ are identified with the boundary coordinates of the manifold $M_m$ (where the membrane theory lives).  This suggests a natural identification between integral curves on $M_m$ and integral curves on $z(x)$.  
Therefore, given a membrane bit thread configuration $w_m$ its bulk bit thread counterpart $w_b$ is simply given by \bea\label{mtob}
w_b^\mu= \Omega \, w_m^\mu\,,
\eea
where $\Omega$ is an undetermined positive function. The hypersurface itself $z(x)$ and the induced metric on $z(x)$, $g_{\alpha \beta}(z(x))$ are determined from the membrane bit thread configuration $w_m^\mu$ via the map (\ref{zu-1}) composed with (\ref{wangle2}).

From our study of the membrane bit threads, we learned that $w_m^\mu$ obeys the bound (\ref{norm-constraint-3}) and it is divergenceless when the background metric of $M_m$ is the flat metric $\delta_{\alpha \beta}$. On the other hand, a given membrane tread configuration $w_m$ give rise to a bulk bit thread configuration $w_b$ together with a background metric $g_{\alpha \beta}(z(x))$ where $w_b$ lives, furthermore we expect $w_b$ on $g_{\alpha \beta}(z(x))$ to obey the same constraints than the bit threads dual to the Ryu-Takayanagi formula obey. We will prove that these constraints can be satisfied by a unique choice of $\Omega$.

\begin{center}
\begin{figure}[!h]
\centering
\includegraphics[scale=0.8]{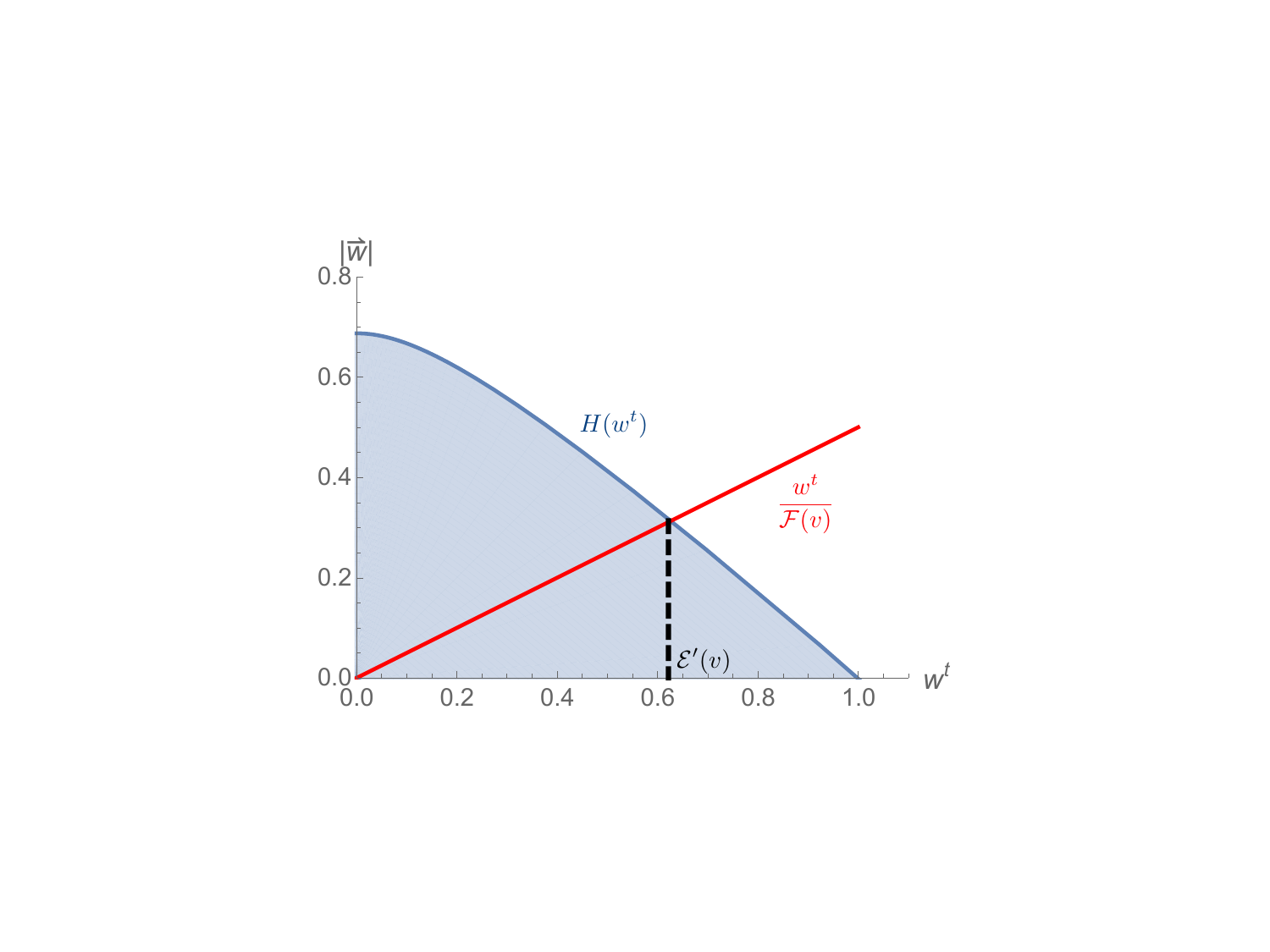}
\caption{We plot the norm bound \eqref{norm-constraint-3} with blue and the relation \eqref{wangle2} with red. To maximize $w^t$ we look for the intersection of the two curves. The blue curve can be parametrized as $(w^t,\abs{\vec{w}})=\le(\mathcal{E}'(\bar{v}),\(\mathcal{E}(\bar{v})-\bar{v}\mathcal{E}'(\bar{v})\)\ri)$ as was explained around \eqref{wsParam}. Recalling the formula for ${\cal F}(v)$ from \eqref{wsAng} we conclude that at the intersection point $\bar{v}=v$ and arrive at \eqref{wmtMax}.
\label{fig:intersection}}
\end{figure}
\end{center}

The bit threads dual to the Ryu Takayanagi formula obey the norm bound 
\bea\label{RT-norm-bound}
g_{\alpha \beta } (z(x)) w_b^{\alpha} w_b^\beta \leq 1\,.
\eea
and are divergenceless with respect to the metric $g_{\alpha \beta}(z(x))$.
The induced metric on the hypersurface \eqref{f-metric-2} in the scaling limit reads 
\bea\label{f-metric-3}
ds^2=\frac{1}{z^{2}}\left( -a(z)dt^{2} +d\vec{x}^{2}  \right)\,,
\eea
Now we plug in this into \eqref{RT-norm-bound} to get 
\es{wSq}{
g_{\alpha \beta } (z(x)) w_b^{\alpha} w_b^\beta&=\frac{\Omega^2}{z^2}\(-a(z)(w_m^t)^2+|\vec{w}_m|^2\)=\frac{(w_m^t)^2\,\Omega^2 }{z^2}\(-a(z)+{1\ov {\cal F}(v)^2}\)\\
&=\frac{(w_m^t)^2\,\Omega^2 }{z^2}\,{\mathcal{E}(v)\(\mathcal{E}(v)-v\mathcal{E}'(v)\)\ov \mathcal{E}'(v)^2}\,.
}
where in the first line we used the definition of $v$ from \eqref{wangle2}, and in the second we replaced $a(z)$ using \eqref{az-1}. Next we want to bound this expression from above, by bounding $(w_m^t)^2$. Since its ratio with $\abs{\vec{w}_m}^2$ is fixed by \eqref{wangle2} and we have the norm bound \eqref{norm-constraint-3}, we are looking for the intersection of the two curves on Fig.~\ref{fig:intersection} to get the maximum of  $(w_m^t)^2$. By what is explained on the figure and its caption, we get 
\es{wmtMax}{
(w_m^t)^2\leq \mathcal{E}'(v)^2\,, \quad \implies \quad g_{\alpha \beta } (z(x)) w_b^{\alpha} w_b^\beta\leq\frac{\Omega^2 }{z^2}\,{\mathcal{E}(v)\(\mathcal{E}(v)-v\mathcal{E}'(v)\)}\,.
}

Comparing \eqref{wmtMax} with (\ref{RT-norm-bound}),  we deduce that  
\bea
\Omega=\frac{z(v)}{\sqrt{\mathcal{E}(v)\(\mathcal{E}(v)-v\mathcal{E}'(v)\)}}\,.
\eea
We observe that $\Omega=1/\sqrt{g}$ by carrying out the following simple computation using \eqref{zu-1} and \eqref{az-1}:
\es{dem}{
\Omega \sqrt{g(z(x))}&=\Omega \,\frac{\sqrt{-a(z)}}{z^d}=\frac{1}{\sqrt{\mathcal{E}(v)\(\mathcal{E}(v)-v\mathcal{E}'(v)\)}}\, \frac{\sqrt{-a(z)}}{z^{d-1}}=1\,, 
}
which leads to the following relationship
\es{wbMap}{
w_b^\mu= \frac{1}{\sqrt{g(z(x))}} \, w_m^\mu\,.
}
This is very convenient, since it immediately connects the divergenceless condition of $w_m^\mu$ with the one of $w_b^\mu$ 
\es{}{
\nabla^{(g)}_\mu w_b^\mu&=\frac{1}{\sqrt{g(z(x))}} \partial_\mu \(\sqrt{g(z(x))}\,w_b^\mu\)=\frac{1}{\sqrt{g(z(x))}} \partial_\mu  w_m^\mu\, \\
&=\frac{1}{\sqrt{g(z(x))}} \nabla^{(\delta)}_\mu  w_m^\mu=0\,,
}
where the superscripts in the covariant derivatives indicates the background metric on which they are defined.

So far we have shown that the bulk bit threads obtained by mapping the membrane bit threads into the bulk according to \eqref{wbMap} obey the constraints $\abs{w_b}_g\leq 1$ and $\nabla^{(g)}_\mu w_b^\mu=0$. Next we show that their flux (as given in \eqref{HRTBitThreadReform}) computes the entanglement entropy $S[A(T)]$.
This then establishes  a precise and consistent map between membrane bit threads and bulk bit threads in the scaling limit.  

Let us start by establishing
\es{fluxEq}{
\sqrt{h}\, w_m^\mu \hn_\mu=\sqrt{h^{(g)}}\, w_b^\mu \hn_\mu^{(g)}\,,
}
where $\sqrt{h^{(g)}}$ is the determinant of induced metric from $g_{\alpha\beta}$ (whereas $\sqrt{h}$ corresponds to the flat metric), while $\hn_\mu^{(g)}$ is the unit normal with respect to $g_{\alpha\beta}$. We want to relate normal covectors defined in different spacetimes as we would like to rewrite  (\ref{flux-membrane}) in terms of a normal covector $\hn^{(g)}_\mu$ defined with respect to the metric $g_{\alpha \beta}( z(x))$. To do so, we start by defining a scalar function $\varphi$ on $M_{m}$ in terms of which we can define the normal covectors via
\bea
\hn_\mu =\frac{\partial_\mu \varphi}{|\partial \varphi|} \,,\qquad \hn^{(g)}_\mu =\frac{\partial_\mu \varphi}{|\partial \varphi|_g}  \,,
\eea
where we use $\abs{\bullet}$ for the norm with respect to the flat metric and $\abs{\bullet}_g$ for the norm with respect to $g_{\alpha \beta}( z(x))$.
Any surface  homologous to $A(T)$ in $M_M$ or on the hypersurface $z(v)$ can be regarded as a level set of a suitable $\varphi$.
Then, we have
\es{NormRel}{
\hn_\mu=\frac{|\partial \varphi|_{g}}{|\partial \varphi|}   \, \hn^{(g)}_\mu \,.
}
The respective induced metric on a given surface with constant $\varphi$ from a background metric $\delta_{\alpha \beta}$ or $g_{\alpha \beta}(z(v))$ is then 
\es{hRel}{
\sqrt{h}= |\partial \varphi|\quad {\rm and}\quad \sqrt{h^{(g)}(v)}=\sqrt{g(z(v))}\, |\partial \varphi|_{g}\,.
}
Putting together \eqref{wbMap}, \eqref{NormRel}, and \eqref{hRel}, we obtain \eqref{fluxEq} that we set out to prove.

\subsection{From bulk bit threads to membrane bit threads}

Let us now rewrite the membrane bit thread program in bulk language. We start with the usual flow formulation, and rewrite the flow maximization in two steps:
\es{m-bit-threads}{
S[A(T)]&=s_{th} \,\underset{w_m}{\rm max} \int_{A(T)}\,\sqrt{h}\, w_m^\mu \hn_\mu\\
&=s_{th} \,\max_{v(x)}\underset{w_m\in {\cal C}[v]}{\rm max} \int_{A(T)}\,\sqrt{h}\, w_m^\mu \hn_\mu\,,
}
where we decomposed the set of vector fields into equivalence classes denoted by ${\cal C}[v]$ that correspond to a given $v(x)$ direction profile according to \eqref{wangle2}, and then performed the maximization in two steps: first we maximize in a fixed equivalence class, then we maximize over the velocity field $v(x)$ characterizing the equivalence class. Note that we omitted the $(g)$ superscript from $h$, this is because on $A(T)$ at $z=1$ there is no $z(v)$ dependence. Using the relation \eqref{fluxEq}, we can rewrite this in bulk language: 
\es{b-bit-threads}{
S[A(T)]
&=s_{th}\, \max_{v(x)}\underset{w_b\in {\cal C}_b[v]}{\rm max} \int_{A(T)}\,\sqrt{h}\, w_b^\mu \hn_\mu^{(g)}\,,
}
where $w_b\in {\cal C}_b[v]$ means:
\es{wangle3}{
{w_b^t(x)\ov \abs{w_b^i(x)}}&={\cal F}(v(x))
}
at every point $x$. Note that the absolute value $ \abs{\vec{w_b}(x)}$ is taken with respect to the flat metric. It is worth noting that, the relation \eqref{wangle3} looks a lot more natural, if we pull down the indices of $w$ with the metric \eqref{f-metric-3}:
\es{wangle4}{
{(w_b)_t\ov \abs{(w_b)_i}}&={g_{tt}(z(v))\ov g_{ii}(z(v))}\,{\cal F}(v)=v\,,
}
where we used \eqref{zu-1}, \eqref{az-1} and the definition of ${\cal F}(v)$ \eqref{wsAng}.

 Finally, using the relation between $v$ and $z$ we can bring this into:
\es{b-bit-threads2}{
S[A(T)]
&={1\ov 4 G_N}\, \max_{z(x)}\underset{w_b\in {\cal C}_b[v(z)]}{\rm max} \int_{A(T)}\,\sqrt{h}\, w_b^\mu \hn_\mu^{(g)}\,.
}
This is very reminiscent of the bit thread dual of the maximin reformulation of HRT \eqref{HRTBitThreadReform}, which we repeat here (with notations modified appropriately):
\es{HRTBitThreadReform2}{
S[A(T)]
&={1\ov 4 G_N} \,\max_{z(x)}\,\underset{w_b}{\rm max} \int_{A(T)}\,\sqrt{h}\, w_b^\mu \hn_\mu^{(g)}\,.
}
We see that the bit thread reformulation of  maximin allows for more general vector fields on the hypersurface defined by $z(x)$, whereas the vector field that we get from mapping the membrane vector flow tie together $z(x)$ and the direction of $w_b(x)$ according to the equation \eqref{wangle3}.

To complete the logic of the paper, we now want to forget the membrane theory and derive the dual program (\ref{m-bit-threads}) directly from the bulk formulation. This is complementary to the work presented in the previous sections, as it might shed light into the way the membrane reinterpretation works. This could be achieved by showing that in \eqref{HRTBitThreadReform2} we can restrict the maximization $\underset{w_b}{\rm max}$ to $\underset{w_b\in {\cal C}[v(z)]}{\rm max}$ and still achieve the same max flow. We do not know how to show this directly in the above formulation.

To complete the derivation, let us rewrite the flux maximization problem over divergenceless bounded vector fields   in an equivalent form, where we make the norm bound explicit by introducing a Lagrange multiplier vector field $\xi$:
\bea
\underset{w}{\rm max} \int_{A(T)}\,\sqrt{h}\, w^\mu \hn_\mu=\underset{\omega}{\rm max} \,\,  \underset{\xi}{\rm min} \left[\int_M \sqrt{g}\,\( |\xi|_g -\xi_\mu \,\omega^\mu \)  +  \int_{A(T)}\,\sqrt{h}\, \omega^\mu \hn_\mu\right]\,,
\eea
where $w$ obeys $\nabla_\mu w^\mu=0$ and $\abs{w}_g\leq1$, while $\nabla_\mu \omega^\mu=0$ but there is no constraint on the norm of $\omega$. Minimizing with respect to $\xi$ leads to the condition $|\omega|_g\leq1$, as desired.

Next, the minimization with respect to $\xi$ and maximization with respect to $\omega$ can be separated into equivalence classes, following a similar story to what we followed above. We write
\es{program-4}{
\underset{w}{\rm max} \int_{A(T)}\,\sqrt{h}\, w^\mu \hn_\mu=\underset{v}{\rm max} \,\, \underset{\omega \in \mathcal{C}_b[v]}{\rm max}\,\, \underset{\bar{v}}{\rm min} \,\, \underset{\xi \in \mathcal{D}_b[\bar{v}]}{\rm min}\, \left[\int_M\sqrt{g}\,\( |\xi|_g -\xi_\mu \,\omega^\mu \)   +  \int_{A(T)}\,\sqrt{h}\, \omega^\mu \hn_\mu\right]\,,
}
 where the relevant sets are:
\es{EquivClasses}{
\mathcal{D}_b[\bar{v}]\equiv\left\{\xi \,\, |\,\,  {\xi_t}=\bar{v}|\xi_i| \right\} \qquad {\rm and }\qquad  \mathcal{C}_b[{v}]\equiv\left\{ \omega \,\, \Big|\,\,{\omega^t}=\mathcal{F}({v}){|{\omega^i}|}\right\}\,,
}
 where for covectors $\mathcal{D}_b[\bar{v}]$ is the analog of  $\mathcal{C}_b[\bar{v}]$ as was explained in \eqref{wangle4}.
 Next, we would like to implement the maximization over the choice of hypersurface parametrized by the function $z(x)$. To proceed,  we have to change variables to:
 \es{omtilde}{
 \tilde \omega^\mu\equiv  \sqrt{g} \, \omega^\mu\,,
 }
 so that only the combination $ \sqrt{g}\,|\xi|_g$ depends on $z(x)$.
 (This holds, because the boundary term does not depend on $z$, since the region $A(T)$ is on the horizon at $z=1$.) The combination  $\tilde \omega^\mu$ is familiar, the same relation relates the membrane and bulk flows in \eqref{wbMap}, but there $z$ was tied to $v$, whereas here it is not (yet). The divergenceless condition $\nabla_\mu \omega^\mu=0$ turns into $\pa_\mu \tilde\omega^\mu=0$, which can be defined independently of the metric. This is the main motivation for the rescaling (\ref{omtilde}).  We have to be aware at this stage that $\tilde\omega^\mu$ does not transform as a vector field under coordinate transformations.
 
 As discussed in Section~\ref{sec:membrane},  in the hydrodynamic limit nothing depends on the derivatives of $z(x)$, then in the maximization step of (\ref{HRTBitThreadReform2}) we can vary with respect to $z(x)$ point by point. We emphasize that domains of  $\xi$ and $\tilde{\omega}$, including the divergencelessness of $\tilde \omega$, are independent of $z(x)$ and then we can carry out the maximization step over $z(x)$ first.  
 We get:
 \es{zvariation}{
 {\pa\ov \pa z}\le( \sqrt{g}\,|\xi|_g\ri)&={\pa\ov \pa z}\le( \sqrt{g(z)}\,\sqrt{g^{tt}(z) \xi_t^2+ g^{xx}(z)\abs{\xi_i}^2}\ri)\\
 &=\abs{\xi_i}{\pa\ov \pa z}\le( {1\ov z^{d-1}}\,\sqrt{ \bar{v}^2-a(z)}\ri)\\
  &=-\abs{\xi_i}{d-1\ov z^{d}\,\sqrt{ \bar{v}^2-a(z)}}\le(  \bar{v}^2-c(z)\ri)\,,
 }
where in the first line we plugged in the definitions, in the second line we used the metric \eqref{f-metric-3}, and in the third we recognized the function $c(z)$ defined in \eqref{cvDef} that ties together $z$ and $\bar{v}$. From \eqref{zvariation} we conclude that $z=z(\bar{v})$, and as expected from the results of Section~\ref{sec:membrane}, we obtain
\es{Ep}{
\sqrt{g}\,|\xi|_g\Big\vert_{z=z(\bar{v})}=\abs{\xi_i}{\cal E}(\bar{v})\,.
}
Thus we can write
\es{program-5}{
\underset{z(x)}{\rm max} \,\, \underset{w}{\rm max} \int_{A(T)}\,\sqrt{h}\, w^\mu \hn_\mu=\underset{v}{\rm max} \,\, \underset{\tilde\omega \in \mathcal{C}[v]}{\rm max}\,\, \underset{\bar{v}}{\rm min} \,\, \underset{\xi \in \mathcal{D}[\bar{v}]}{\rm min}\, \left[\int_M \( \abs{\xi_i}{\cal E}(\bar{v})-\xi_\mu \tilde\omega^\mu \)  +  \int_{A(T)}\,\sqrt{h}\, \tilde\omega^\mu \hn_\mu\right]\,, 
}
where we used the fact that on the boundary $\omega=\tilde\omega$. On the RHS we recognize the membrane bit thread problem, if we rewrite the two step minimization and maximization steps as $\max_{\tilde \omega} \min_\xi$, and compare to \eqref{minLag} with identification $\tilde \omega=w,\, \xi=-N$, and the Lagrange multipliers $\phi,\, \psi$ in \eqref{minLag} eliminated making the divergenceless constraint implicit and restricting the integration of the flux through $A(T)$. This completes the goal we set out to achieve: we derived the membrane bit thread problem starting from bulk bit threads without referring to maximin surfaces or membranes. 

We deepen our understanding of the mapping of bulk bit threads to membrane bit threads in the next section.

\subsection{Tying the hypersurface to the bit thread direction}

It is worth keeping more metric information in the membrane bit thread problem to gain further insight. In the membrane bit thread problem given in \eqref{flux-membrane}, the only remnant of the bulk metric is through the nontrivial norm bound \eqref{norm-constraint-3}. Instead it would be nice to recover the formulation \eqref{b-bit-threads2} directly. 

To do this, we 
 perform the minimization over $\xi$ in the form $ \underset{\bar{v}}{\rm min} \,\, \underset{\xi \in \mathcal{D}[\bar{v}]}{\rm min}\,\bullet$.
Since we want to keep the notion of a metric in our final program, we need to associate the minimization over $\bar{v}$ with the appropriate angular variable in the variation over $\xi_\mu$. In Section~\ref{sec:vMeaning} in \eqref{wsParam} we made one choice in parametrizing a vector field with $\bar{v}$ and $\lambda$, here we will make another one. Let us define $\lambda$ by:
\bea
\lambda\equiv \sqrt{g}\,|\xi|_g\Big\vert_{z=z(\bar{v})}=\abs{\xi_i}{\cal E}(\bar{v})\geq0\,,
\eea
from which the parametrization for $\xi_\mu$ is
\bea
\xi_\mu=\(\xi_t, \xi_i \)={\lambda\ov {\cal E}(\bar{v})} \( \bar{v}, n_i\)\,,
\eea
with $n_i=\xi_i/|\xi_i|$.\footnote{A possible other parametrization would be to keep $|\xi_i|=\text{const.}$, but since $\xi_t=\bar{v}|\xi_i|$, varying $\bar{v}$ would mean varying   $\xi_t$, and  $\bar{v}$ would not be an angular variable.} Then the first term in \eqref{program-5} is
\bea \label{bound-constraint-2}
\int_M \( \abs{\xi_i}{\cal E}(\bar{v})-\xi_\mu \tilde\omega^\mu \) =\int_M \lambda\(1- \frac{1}{{\cal E}(\bar{v})}\(\bar{v}\tilde{\om}^t+n_i {\tilde{\om}^i} \) \)\,.
\eea
Minimizing over $\bar{v}$ and $n_i$ leads to the solutions 
\bea
n_i=\delta_{ij} \frac{\tilde{\om}^j}{|\tilde{\om}^k|}, \quad  {\rm and} \quad  {\cal F}(\bar{v}) ={\tilde\omega^t\ov \abs{\tilde \omega^k}} \,.
\eea
Comparing to \eqref{EquivClasses}, we conclude that $v=\bar{v}$. Plugging this relation back into (\ref{bound-constraint-2}) and imposing  positivity  leads to the bound on $\tilde{\om}$:
\bea\label{norm-bound-4}
\(1- \frac{1}{{\cal E}(v)}\(v\tilde{\om}^t+|{\tilde{\om}^i}| \) \)
&=&\(1- \frac{\tilde{\om}^t}{{\cal E}(v)}\(v +\frac{1}{\mathcal{F}(v)} \) \)\geq 0\,.
\eea
Using the definition of ${\cal F}$ in \eqref{wsAng}, the above simplifies to $\tilde{\om}^t \leq  {\cal E}'(v)$, thus $|\tilde{\om}^i| \leq {\cal E}(v)- v {\cal E}'(v)$ from \eqref{EquivClasses}. 
We can explicitly verify that this leads to the expected norm bound  after we rescale back to the original variable  $\omega$, that is
 \es{OutCome}{
\abs{ \omega}_{g(v)}&={\abs{\tilde \omega}_{g(v)}\ov \sqrt{g(z(v))}}=\sqrt{z(v)^{2(d-1)}\le(\le(\tilde\omega^t\ri)^2+{\abs{\tilde \omega^i}^2\ov -a(z(v))}\ri)}\\
&=\abs{\tilde\omega^t}\sqrt{z(v)^{2(d-1)}\le(1+{1\ov -a(z(v))\, {\cal F}(v)^2}\ri)} \\
&= \frac{\abs{\tilde\omega^t}}{| {\cal E}'(v)|}\,\, {\leq} \, 1 \,,
}
where in the second line we used the relations \eqref{zu-1} and \eqref{az-1},  and  in the third line  we used the bound derived from (\ref{norm-bound-4}). The last step is to evaluate the minimum of the bulk term in (\ref{program-5}) which equals zero at $\lambda=0$. 

We conclude that
\bea
\underset{z(x)}{\rm max} \,\, \underset{w}{\rm max} \int_{A(T)}\,\sqrt{h}\, w^\mu \hn_\mu=\underset{v}{\rm max} \,\, \underset{\omega \in \mathcal{C}[v]}{\rm max}\,  \int_{A(T)}\,\sqrt{h}\, \omega^\mu \hn_\mu\,,
\eea
with the implicit constraints on the LHS $\nabla_\mu w^\mu=0$ and $\abs{w}_g\leq 1$, while on the RHS $\nabla^{(g(v))}_\mu \omega^\mu=0$ and $\abs{\omega}_{g(v)}\leq 1$. The LHS is the bulk bit thread problem, while the RHS is the rewriting of the membrane bit thread problem that we arrived at in \eqref{b-bit-threads2}. This concludes the alternative derivation of the membrane flow program from the bulk flow program in the hydrodynamic limit.

\section{Discussion}
In the present work we have succeeded in obtaining a flow based program dual to the membrane theory of entanglement dynamics. We derived such a program from two complementary starting points: one was the membrane theory itself and the other one, was the max flow program dual to the HRT formula. For the first derivation we used convex optimization techniques to dualize the min cut program (\ref{SA-2}), which was shown to be equivalent to the membrane theory problem (\ref{MembraneTheory2}). The second derivation uses the flow program dual to the HRT formula, RHS of (\ref{HRTBitThreadReform}) and by studying its hydrodynamic regime of large times and large volumes, we arrived at the same dual formulation (\ref{MemBitThreadReform}). These two independent derivations serve as a consistency check of our main result (\ref{MemBitThreadReform}), and shows the mutual consistency of the hydrodynamic limit and the bit thread reformulations of the holographic entanglement entropy formula as illustrated in figure \ref{fig:flowchart}. Additionally, the above results help to clarify the relation between membrane and HRT bit threads.

Having this dual formulation for the membrane theory serves two important roles. First, within the context of holographic entanglement, our flow reformulation (\ref{MemBitThreadReform}) represents the first example of a max flow program dual to the holographic entanglement entropy of a fully dynamical situation that has appeared in the literature.  Second, our reformulation (\ref{MemBitThreadReform}) takes bit threads from the realm of holography to the domain of generic quantum systems:
we emphasize that the membrane theory is proposed to describe the entanglement dynamics of generic chaotic systems (and was explicitly derived both for holographic field theories \cite{Mezei:2016wfz,Mezei:2016zxg,Mezei:2018jco} and for random quantum circuits \cite{Nahum:2016muy,Jonay:2018yei}).
By interpreting the entanglement entropy of a region $A$ as given by the maximum number of one dimensional threads connecting the region with its complement -- thought of them as representing EPR like entanglement -- our formula makes manifest the quantum information theoretic meanings of properties such as subadditivity and strong subadditivity, in the exact same way as it happened with the holographic entanglement entropy formula (in static setups) \cite{Freedman:2016zud}. 
It would be very interesting if this new reformulation of the membrane theory could serve as a way to prove the hypothesized validity of the membrane theory for all quantum chaotic systems. 

\subsection*{Acknowledgments}
We are especially indebted to Matthew Headrick for multiple discussions on convex optimization and differential geometry, and for his invaluable feedback on preliminary versions of our work. We would also like to thank Jan De Boer, Netta Engelhardt, Juan Pedraza and Julio Virrueta for useful discussions. CAA is supported by the National
Science Foundation under CAREER award PHY16-20628. MM is supported by the Simons Center for Geometry and Physics.

\appendix

\section{Comments on maximin}\label{app:maximin}

\begin{figure}
\centering
 \begin{overpic}[width=2.7in
 ]{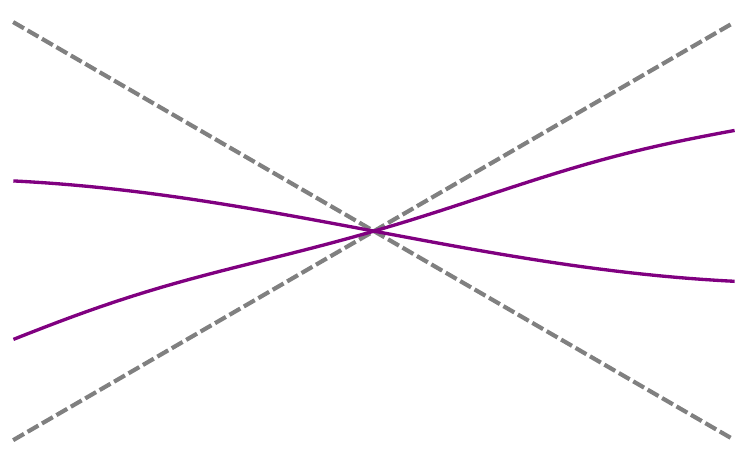}
 \put (85,38) {\footnotesize{$\Sigma_2$}}
  \put (87,20) {\footnotesize{$\Sigma_1$}}
  \put (48.6,30) {\scriptsize{$\bullet$}}
   \put (44,24) {\footnotesize{$M(A)$}}
 \end{overpic}
  \begin{overpic}[width=2.7in
 ]{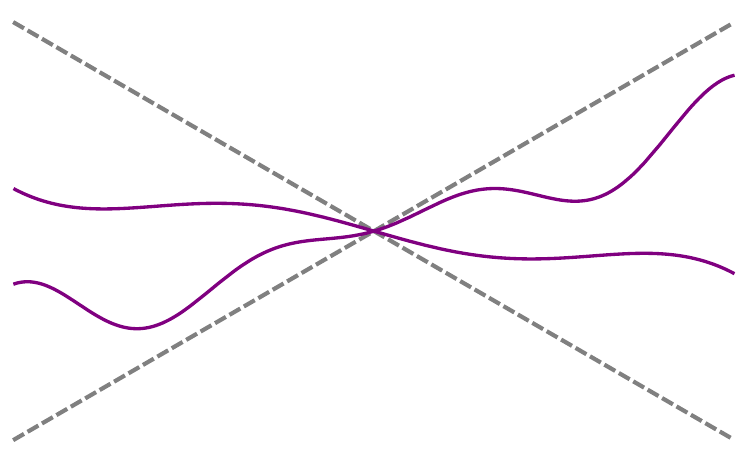}
 \put (87,38) {\footnotesize{$\Sigma_{\phi_2}$}}
  \put (87,23) {\footnotesize{$\Sigma_{\phi_1}$}}
  \put (48.6,30) {\scriptsize{$\bullet$}}
   \put (43,24) {\footnotesize{$m_{\!_M}\!(A)$}}
 \end{overpic}
 \caption{Schematic representation of the cone like regions defined around the maximin $M(A)$ and minmax $m_M(A)$ surfaces on the left and on the right respectively. On the left, we illustrate two hypersurfaces $\Sigma_1$, $\Sigma_2$ where $M(A)$ is a minimal area surface homologous to $A$. On the right we illustrate two hypersurfaces $\Sigma_{\phi_1}$, $\Sigma_{\phi_2}$ on which $m_M(A)$ is a minimal area surface among the induced level set surfaces of $\phi_1$ on $\Sigma_{\phi_1}$ and $\phi_2$ on  $\Sigma_{\phi_2}$ respectively. \label{maximin-minmax}}
\end{figure}
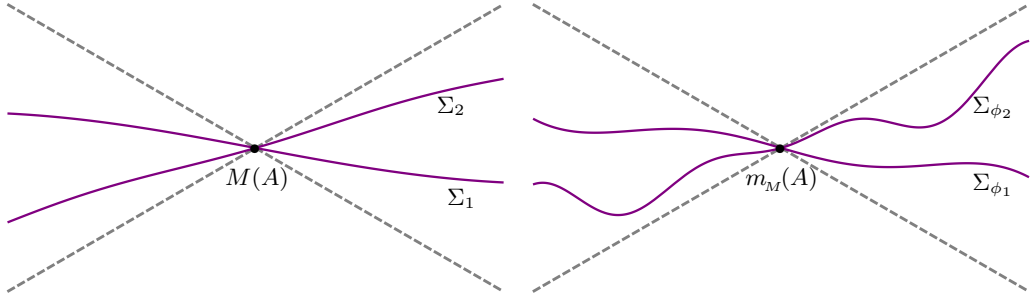

We present a compelling argument for the validity of the equality between (\ref{maximin2}) and (\ref{maximin3}), assuming that both maximin $M(A)$ and minmax $m_M(A)$ surfaces are unique. Typically, a solution of (\ref{maximin2}) gives the maximin surface $M(A)$ and a particular choice of hypersurface $\Sigma$, say $z_M(x)$ on which one achieved maximality of area$[m(A)]$  under variations along $z$. However, $\Sigma$ is highly nonunique away from $M(A)$ \cite{Wall:2012uf}, and therefore, the set of $\Sigma$ on which $M(A)$ is a minimal area surface homologous to $A$ lies inside a cone-like bulk region which converges on $M(A)$, as illustrated in figure \ref{maximin-minmax}. Now, let us consider a scalar function $\phi(x)$ defined on $\Sigma_0$ such that its level set surfaces are homologous to $A$, with its zero level set surface being equal to $m_M(A)$. This function will define a hypersurface $\Sigma_{\phi}$, obtained via the area maximization of each of its level set surfaces with respect to variations along the $z$ direction, step one of (\ref{maximin3}). A set of functions $\phi_{\alpha}(x)$ which satisfy the above properties give rise to a set of hypersurfaces $\Sigma_{\phi_\alpha}$ which lie inside a cone-like bulk region that converges on $m_M(A)$, as illustrated in figure \ref{maximin-minmax}. For each $\Sigma_\phi$, the surface $m_M(A)$ has minimal area among the set of surfaces on $\Sigma_\phi$ induced from the level set surfaces of $\phi(x)$.  Since both $M(A)$ and $m_M(A)$ are maximal under variations along $z$, lies at the intersection of infinitely many hypersurfaces $\Sigma_\alpha$ and $\Sigma_{\phi_\alpha}$, respectively, and their areas are minimal with respect to any family of surfaces homologous to $A$ on the appropriate member of these hypersurfaces, then, these surfaces must coincide with each other, $M(A)=m_M(A)$.

\section{Mapping between the bulk and the membrane pictures\label{B}}
Let us start by recalling the main equations from Section~\ref{section2}. The first is 
\bea \label{v^2}
v^2=a(z)-\frac{z a'(z)}{2(d-1)}\equiv c(z)\,,
\eea
where $v$ encodes the angle of the normal of the membrane with the time direction.
The second is the membrane tension function $\mathcal{E}(v)$ given by
\bea \label{epsilon}
\mathcal{E}(v)= \sqrt{ \frac{-a'(z)}{2(d-1)z^{2d-3}}}\Bigg|_{z=c^{-1}(v^2)}\,.
\eea
We want to obtain a formula for $z(v)$ starting from the knowledge of $\mathcal{E}(v)$. Implicit differentiation of (\ref{v^2}) and (\ref{epsilon}) leads to the following equations
\es{pair}{
2v\frac{dv}{dz}&=\frac{(2d-3)a'(z)-z a''(z)}{2(d-1)}\,,  \\
2 \mathcal{E}(v) \mathcal{E}'(v) \frac{dv}{dz}&=\frac{(2d-3)a'(z)-z a''(z)}{2(d-1) z^{2d-2}} \,.
}
The second equation is obtained by first taking the square of  (\ref{epsilon}) and then implicitly differentiating with respect to $z$. Once in this form, one finds immediately a closed expression for $z(v)$,
\bea\label{z(v)}
z(v)=\(\frac{v}{\mathcal{E}(v) \mathcal{E}'(v)}\)^{\frac{1}{2(d-1)}}\,.
\eea
Notice that the reason why one can solve for $z$ is simply because in both equations (\ref{pair}) one finds precisely the same combination of $a'(z)$ and $a''(z)$. Then in their ratio one is left with a function of $v$ and $z$ to some power.

Now to obtain $a(z(v))$ we start from (\ref{v^2}), express $a'(z)$ by a combination of $\mathcal{E}(v)$ and $z$ from (\ref{epsilon}), which then leads to 
\bea
a(z)=v^2-z^{2(d-1)}  \mathcal{E}^2(v)\,.
\eea
Since we already have $z$ as a function of $v$ in (\ref{z(v)}), then this reduces to 
\bea\label{a(z)}
a(z)=-v\(\frac{\mathcal{E}(v) -v\mathcal{E}'(v) }{\mathcal{E}'(v) }\)\,.
\eea

\bibliographystyle{ucsd}
\bibliography{refs-flows}

\end{document}